\documentclass[]{jfm} 

\usepackage{multirow}
\usepackage{makecell}

\usepackage{graphicx}
\usepackage{newtxtext}
\usepackage{newtxmath}
\usepackage{natbib}
\usepackage{hyperref}
\hypersetup{
    colorlinks = true,
    urlcolor   = blue,
    citecolor  = black,
}

\usepackage{overpic}[grid=True]
\newcommand{\RomanNumeralCaps}[1]
\linenumbers

\usepackage{siunitx}
\usepackage[dvipsnames]{xcolor}

\newcommand{\ReT}{\mathrm{Re}_\mathrm{T}}
\newcommand{\dan}[1]{{#1}} 
\newcommand{\dann}[1]{{#1}} 
\newcommand{\new}[1]{{#1}} 
\renewcommand\vec{\boldsymbol}

\title{Structure and energy transfer in homogeneous turbulence below a free surface}

\author{Daniel J. Ruth\aff{1}
  \corresp{\email{daruth@ethz.ch}},
  Filippo Coletti\aff{1}}

\affiliation{\aff{1}Institute for Fluid Dynamics, ETH Zurich, Zurich, Switzerland}

\begin{document}
\maketitle

\begin{abstract}
We investigate the turbulence below a quasi-flat free surface, focusing on the energy transport in space and across scales. We leverage a large zero-mean-flow tank where homogeneous turbulence is generated by randomly actuated jets. A wide range of Reynolds number is spanned, reaching sufficient scale separation for the emergence of an inertial sub-range. Unlike previous studies, the forcing extends through the source layer, though the surface deformation remains millimetric. Particle image velocimetry along a surface-normal plane resolves from the dissipative to the integral scales. \new{The contributions to turbulent kinetic energy from both} vertical and horizontal components \new{of velocity} approach the prediction based on rapid distortion theory as the Reynolds number is increased, indicating that discrepancies among previous studies are likely due to differences in the forcing. At odds with the theory, however, the integral scale of the horizontal fluctuations grows as the surface is approached. This is rooted in the profound influence exerted by the surface on the inter-scale energy transfer: along horizontal separations, the direct cascade of \new{energy in horizontal fluctuations} is hindered, while an inverse cascade of \new{that in vertical fluctuations} is established. This is connected to the structure of upwellings and downwellings. The former, characterized by somewhat larger spatial extent and stronger intensity, are associated to extensional surface-parallel motions. They thus transfer energy to the larger horizontal scales, prevailing over downwellings which favour the compression (and concurrent vertical stretching) of the eddies. Both types of structures extend to depths between the integral and Taylor microscales.
\end{abstract}


\section{Introduction}
\label{sec:introduction}

Turbulent liquid flows often involve a free surface as an upper boundary; consider, for instance, the ocean upper layer separated from the atmosphere by the air-sea interface, or the surface in liquid mixing vessels used in \new{many} industrial processes. To understand the flow physics common to such situations, it is useful to consider the archetypical case in which the free surface bounds an otherwise homogeneous and isotropic region of zero-mean flow turbulence. While this has been extensively investigated, our understating of this fundamental and highly relevant class of flows is still incomplete. With no ambition to provide a full account of the literature, below we briefly describe the problem, summarize the picture painted by some key studies, and single out important open questions that motivate the present work.

\subsection {Description of the problem}

So long as gravity or surface tensions keeps the deformation of the surface to a minimum, the surface-normal (vertical) motions vanish approaching the surface. For this reason, many aspects of the situation resemble zero-mean-flow turbulence adjacent to a solid boundary \citep{perot_shear-free_1995}. Unlike a solid wall, however, a clean free surface imposes a shear-free boundary condition at the surface, which allows surface-parallel (horizontal) velocities to persist. In their hallmark study, \citet{hunt_free-stream_1978} invoked rapid distortion theory (RDT) to predict the behaviour of an otherwise homogeneous isotropic turbulent flow adjacent to a flat plate. Their analysis, as well as several successive studies (e.g., \citet{brumley_near-surface_1987}; \citet{shen_surface_1999}; \citet{teixeira_distortion_2002}; \citet{magnaudet_high-reynolds-number_2003}) distinguished between two layers beneath the surface $z=0$ (where $z$ is the vertical upward coordinate).

The so-called source layer or blockage layer, extending to a depth $z\sim-L_\infty$ (where $L_\infty$ is the integral scale of the turbulence \new{far from the surface}) represents the region in which the kinematic (no-penetration) boundary condition is felt. In this region, the \new{contribution to turbulent kinetic energy (TKE) from vertical velocity fluctuations}, $\overline{u_z^2}$, with overlines denoting averages in time, decays to zero. As upwards-moving fluid travels towards the surface through the source layer (upwellings or splats), the no-permeability condition induces an inter-component transfer of energy from vertical to horizontal motions and the \new{contribution to TKE from horizontal velocity fluctuations}, $\overline{u_x^2}$, is enhanced. This energy is partly transferred back to vertical TKE when regions of surface-tangential flow converge and are redirected downwards (downwellings or anti-splats); see \citet{perot_shear-free_1995}. \new{While TKE $\overline{\vec{u} \cdot \vec{u}}$ is a scalar quantity, for simplicity from hereon we refer to $\overline{u_x^2} \approx \overline{u_y^2}$ and $\overline{u_z^2}$ as horizontal and vertical TKE, respectively, and the re-partioning of energy between $\overline{u_x^2}$ and $\overline{u_z^2}$ as inter-component energy transfer.}

The dynamic boundary condition affects a shallower viscous layer, $z>-\dann{\delta_\nu} \new{=} L_\infty \ReT^{-1/2}$, where the velocity gradients are modified to satisfy the zero-shear-stress condition at the surface. The problem is parametrized with the bulk Reynolds number
\begin{equation}
    \ReT= \frac{2 u'_\infty L_\infty}{\nu}, \label{eq:ReT}
\end{equation}
where $\nu$ is the fluid kinematic viscosity. Here and in the following, the prime indicates the root mean square (r.m.s.) of the fluctuations around the mean and the subscript $\infty$ indicates quantities averaged over the homogeneous bulk. With a purely flat surface and shear-free interface, $\mathrm{Re}_\mathrm{T}$ fully defines the problem when the turbulence in the bulk is homogeneous and isotropic and its spatial decay (absent surface-induced effects) is negligible. In practice, the surface is deformable to the extent that gravity and surface tension cannot suppress turbulent fluctuations. We will focus on regimes in which the effect of surface deformation on the flow is small.

\subsection{Previous studies}

Early experiments investigated the interaction of turbulence with a solid boundary imposing no mean shear on the flow. \citet{uzkan_shear-free_1967} and \citet{thomas_grid_1977} considered grid turbulence interacting with a flow-parallel wall traveling at the fluid’s mean velocity, finding an increase in the horizontal TKE at the expense of the vertical TKE near the surface. More recently, \citet{johnson_turbulent_2018} investigated zero-mean-flow turbulence generated by a randomly-actuated jet array opposite a solid wall, finding similar behaviour of the TKE \new{partitioning}. Those studies found that the depth of the layer influenced by the surface was $O(L_\infty)$, in agreement with the prediction of \citet{hunt_free-stream_1978}.

Seminal experiments on zero-mean flow turbulence below a free surface were conducted by \citet{brumley_near-surface_1987} up to $\mathrm{Re}_\mathrm{T} \sim 370$, who used an \dann{oscillating} grid and observed an increase in horizontal TKE at the expense of vertical TKE as the surface was approached, in agreement with \citet{hunt_free-stream_1978}. Similar results were reported at much larger $\mathrm{Re}_\mathrm{T}$ by \citet{variano_turbulent_2013} using \dann{a random-jet-array system similar to that of \citet{johnson_turbulent_2018}}. They \dann{additionally} found a decrease of horizontal TKE just beneath the surface, which was attributed to unavoidable surface contamination by surfactants, inhibiting surface dilatational motions \citep{shen_effect_2004}. \citet{herlina_experiments_2008}, on the other hand, did not observe an increase in horizontal TKE, and attributed the disagreement with \citet{hunt_free-stream_1978}’s theory to its simplifying assumptions, in particular its inviscid nature.

Mechanisms controlling the TKE budget were analysed in the numerical study by \citet{perot_shear-free_1995}, who considered various types of boundary conditions. Comparison with a solid wall boundary suggested that the extent of inter-component transfer of energy is due to the imbalance between up- and downwellings. Their simulations, as well as those by \citet{guo_interaction_2010} and \citet{herlina_simulation_2019}, suggested that upwellings are more energetic than downwellings, pointing to an important role of their imbalance in determining the free-surface flow dynamics. Numerical simulations by \citet{walker_shear-free_1996} and \citet{teixeira_distortion_2002} highlighted how the dynamic boundary condition induces a smaller dissipation rate at the surface, while it does not significantly alter the surface-normal vorticity.

\subsection{Open questions and motivation for the present study}

The applicability of the \citet{hunt_free-stream_1978}'s theory to sub-surface turbulence was debated in several experimental, numerical and theoretical studies, as reviewed in \citet{magnaudet_high-reynolds-number_2003}. While there is substantial evidence that such theory is in qualitative agreement with the observations, quantitative comparisons have been limited, in particular concerning its predictions on the gradients and correlation scales in the near-surface region. Verification of the theory has been complicated by the way sub-surface turbulence is introduced. In some configurations, this is forced several integral length scales away from the free surface, and any effect of the latter is superimposed on the spatial decay of the turbulence (e.g.; \citet{walker_shear-free_1996}). In others, homogeneous turbulence is generated as an initial condition before the surface is suddenly introduced, yielding an inherently transient behaviour \citep{perot_shear-free_1995}. Moreover, as RDT is essentially inviscid, its predictions are expected to apply in the limit of high $\mathrm{Re}_\mathrm{T}$. Systematic studies of Reynolds number effects, however, have not been conducted.

The presence of the surface profoundly transforms the nature of the turbulence in its immediate proximity. \new{Already, \citet{eckhardt_turbulence_2001} showed with numerical simulations that intermittancy in the sub-surface velocity increases near the surface. \citet{cressman_eulerian_2004} paired these findings with experimental data showing that floating tracers disperse less rapidly than is the case in sub-surface turbulence, attributing this to the two-dimensionality imposed by the free-surface boundary condition. \citet{perot_shear-free_1995} similarly suggested that, along the surface, the two-dimensional (2D) nature of the flow alters the direct energy cascade expected in three-dimensional (3D) flows.} While this view was supported by \new{simulations of} open channel flows \citep{pan_numerical_1995,lovecchio_upscale_2015}, the majority of studies on homogeneous turbulence under a free surface argued that the flow is essentially 3D, in that the boundary condition does not impede vortex stretching and the associated down-scale energy transfer \citep{walker_shear-free_1996,shen_surface_1999,guo_interaction_2010}.

As mentioned, the complex sub-surface dynamics are heavily influenced by the balance between upwellings and downwellings. These act as building blocks of the near-surface flow, and their properties are critical to the renewal of the free surface (and thus the associated gas transfer) \citep{guo_interaction_2010,kermani_surface_2009,variano_turbulent_2013,herlina_direct_2014}. Gas transfer rates have been directly linked to the free-surface divergence $\beta=-\partial u_z / \partial z$ (with the velocity gradient evaluated at $z = 0$), whose sign and magnitude depends on the upwelling/downwelling state of the sub-surface flow \citep{jahne_air-water_1998,mckenna_role_2004,turney_airwater_2013}. \new{In particular, recent work \citep{babiker_vortex_2023} suggests that near-surface mixing can be quantified by observations of minuscule deformations to the surface; such an approach, clearly,  requires an understanding of the connection between the characteristics of the surface divergence evidenced by the deformations and the turbulence. However, the} spatial and velocity scales of upward and downward motions have been examined mostly in numerical studies at limited $\mathrm{Re}_\mathrm{T}$, and therefore their extent and strength in regimes relevant to environmental and industrial settings have not been established. 

Motivated by these considerations, here we analyse the results of an extensive measurement campaign focused on the effects of a quasi-flat free surface on an otherwise homogeneous turbulent flow. Unlike previous studies, we consider a system in which high-$\ReT$ turbulence is steadily forced in the vicinity of the surface, minimizing spatial variations unrelated to the effect of the surface. By means of high-resolution particle image velocimetry (PIV) and laser-induced fluorescence (LIF), we characterize the turbulence structure from the bulk region to the free surface, resolving from the dissipative to the integral scales of the flow. The paper is organized as follows. In section \ref{sec:experimental}, we present the experimental facility, the imaging methodology, and the flow statistics that define the regime under consideration. In section \ref{sec:turbulence_modulation}, we analyse the structure and evolution of the turbulence \new{between the bulk and the surface}, systematically comparing our observations with RDT predictions and exploring the inter-scale energy transfer. In section \ref{sec:upwellings_downwellings}, we focus on the respective roles of upwellings and downwellings in the transport of energy in space and across scales. We summarize the main findings and draw conclusions in section \ref{sec:conclusions}.

\section{Experimental methodology and flow regime}
\label{sec:experimental}

\subsection{Apparatus and measurement approach}

\begin{figure}
  \centerline{\begin{overpic}[width=\linewidth]{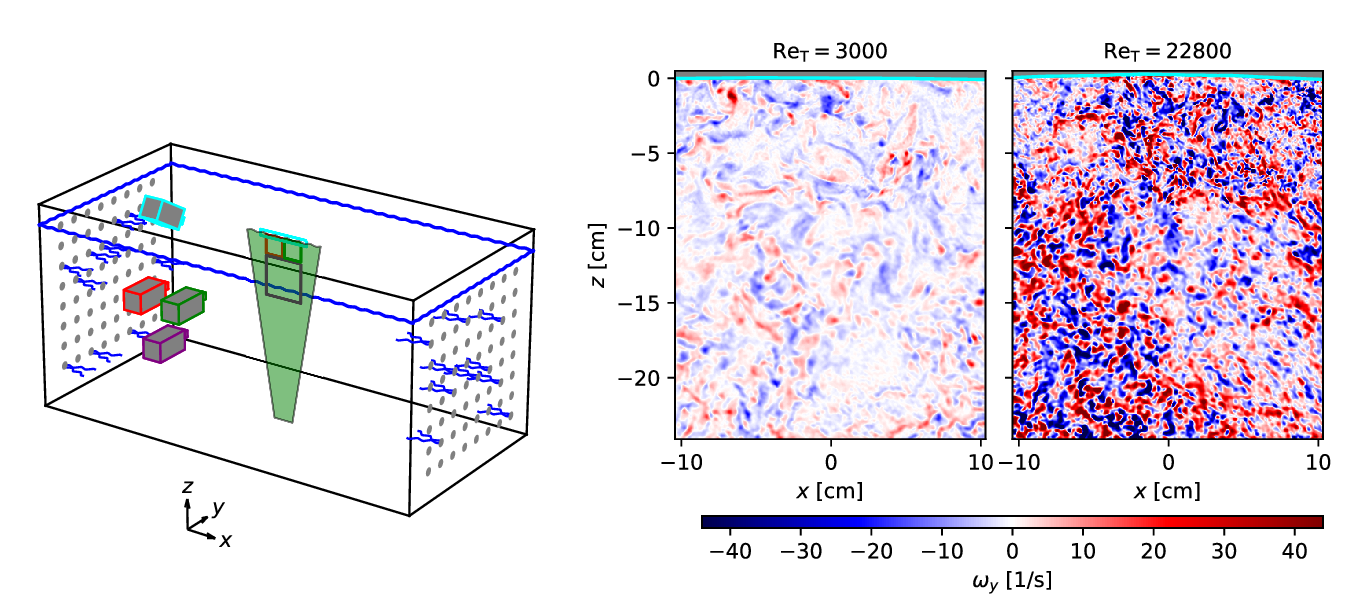}\put(0,30){(a)}\put(50,40){(b)}\put(75,40){(c)}\end{overpic}}
  \caption{(a) The turbulence tank and imaging system. Pumps on either side of the tank emit jets of water (blue) randomly. The three PIV cameras colored red, green, and purple each resolve the FOVs indicated within the laser sheet (green). A fourth camera (colored cyan) resolves the surface position using laser-induced fluorescence. (b-c) Snapshots of the out-of-plane vorticity fields at the lowest and highest Reynolds numbers, respectively. As the Reynolds number increases, the magnitude of the vorticity increases and its spatial scale decreases. }
\label{fig:experiment_schematic}
\end{figure}

The experimental apparatus is illustrated in figure \ref{fig:experiment_schematic} (a). Turbulence is created in a \SI{2}{m^3} water tank by two opposing $8 \times 8$ arrays of submerged pumps. Within each array, the pumps are separated by \SI{10}{cm} in the horizontal and vertical directions and intermittently emit turbulent jets according to the “sunbathing” algorithm proposed by \citet{variano_turbulent_2013}. The magnitude of the fluctuating velocity, and consequently the bulk Reynolds number $\ReT$, is changed by modulating the power supplied to each pump. This is controlled by programmable logic circuits, dictating a pulse-width-modulation scheme for each pump \citep{chan_settling_2021}. On average, 12.5\% of pumps are turned on at a given time and each jet emission lasts \SI{3}{s}. The water level is approximately \SI{5}{cm} above the axis of the jets in the top row of the array. The relatively small distance between the forcing region and the surface distinguishes the present setup from the majority of previous experimental efforts, which employed oscillating grids or actuated jets placed several integral scales below the surface (e.g., \citet{brumley_near-surface_1987,mckenna_role_2004,herlina_experiments_2008,variano_random-jet-stirred_2008,variano_turbulent_2013}). \citet{savelsberg_experiments_2009} also forced turbulence close to the surface with an active grid in an open channel flow, but did not investigate the influence of the surface on the turbulence underneath. The surface tension of the water $\sigma$ is measured via a Du Noüy ring at various points in time, yielding no significant variations around the standard value of \SI{0.07}{N/m}.

The velocity field in the centre of the tank is measured by particle image velocimetry (PIV). A \SI{532}{nm} laser beam (Nd:YAG, \SI{200}{mJ/pulse}) is converted into a thin diverging sheet and shone vertically through the glass bottom surface of the tank, illuminating a region within the plane $y=0$ (see figure \ref{fig:experiment_schematic} (a)). We denote with $x$ the horizontal direction parallel to the jet axes, and $z$ the vertical upwards direction, with the origin at the water surface. As is sketched in figure \ref{fig:experiment_schematic} (a), three synchronized cameras (CMOS, 25 Megapixels) are used to image two side-by-side regions just below the surface, as well as a larger region beneath. The tracers are \SI{10}{\micro m} hollow glass sphere particles, and the inter-frame timing is varied with $\ReT$ to ensure their maximum typical displacement is approximately 5 pixels, optimal for zero-mean-flow turbulence facilities of this kind \citep{carter_generating_2016}. In total, the field of view (FOV) resolved by the cameras extends approximately \SI{20}{cm} in the horizontal direction and approximately \SI{25}{cm} below the free surface, centred on the midpoint between the two arrays of pumps. Between 4000 and 6000 instantaneous velocity fields are obtained for each condition at a rate of \SI{1}{Hz} using an iterative cross-correlation algorithm \citep{thielicke_pivlab_2014}. Velocity components $U_x$ and $U_z$ from the \new{three} cameras are interpolated onto a uniform grid \new{with spacing \SI{0.53}{mm}, which is comparable to the vector spacing obtained with the higher-resolution cameras resolving the region just beneath the surface. Such vector spacing is the result of an interrogation window size of about \SI{1}{mm} (which defines the PIV spatial resolution) followed by a standard 50\% overlap between adjacent windows. The PIV resolution deeper into the bulk is coarser, with an interrogation window size of approximately \SI{2}{mm}.}

A small amount (less than \num{3e-7} in volume) of uranine dye is added to the water to capture the instantaneous position of the water surface $\eta(x)$ by LIF. To this end, a fourth CMOS 25 megapixel camera synchronized with the laser pulse is outfitted with a band-stop filter to block the bright laser light and positioned above the water surface, angled down by approximately \SI{30}{degrees}. It captures the fluorescence of the dye, with the uppermost part of the bright region demarking the water surface position.

Snapshots of the vorticity field and surface position are shown in figure \ref{fig:experiment_schematic} (b) and (c) at the lowest and highest Reynolds numbers investigated, respectively, highlighting the finer structures at the higher turbulence \dann{intensity}. Animations of the vorticity fields from each case (recorded at a faster frame rate for the purpose of visualization) are provided as a supplementary video.

\subsection{Turbulence properties in the bulk}

The turbulence statistics are impacted by the presence of the free surface within approximately one bulk longitudinal integral scale $L_\infty$ from the free surface \citep{hunt_free-stream_1978}. As described below, $L_\infty \approx \SI{10}{cm}$; as such, in this section we show results spatially averaged over $z< \SI{-15}{cm}$, where the flow statistics vary marginally with depth.

\begin{figure}
  \centerline{\begin{overpic}[width=\linewidth]{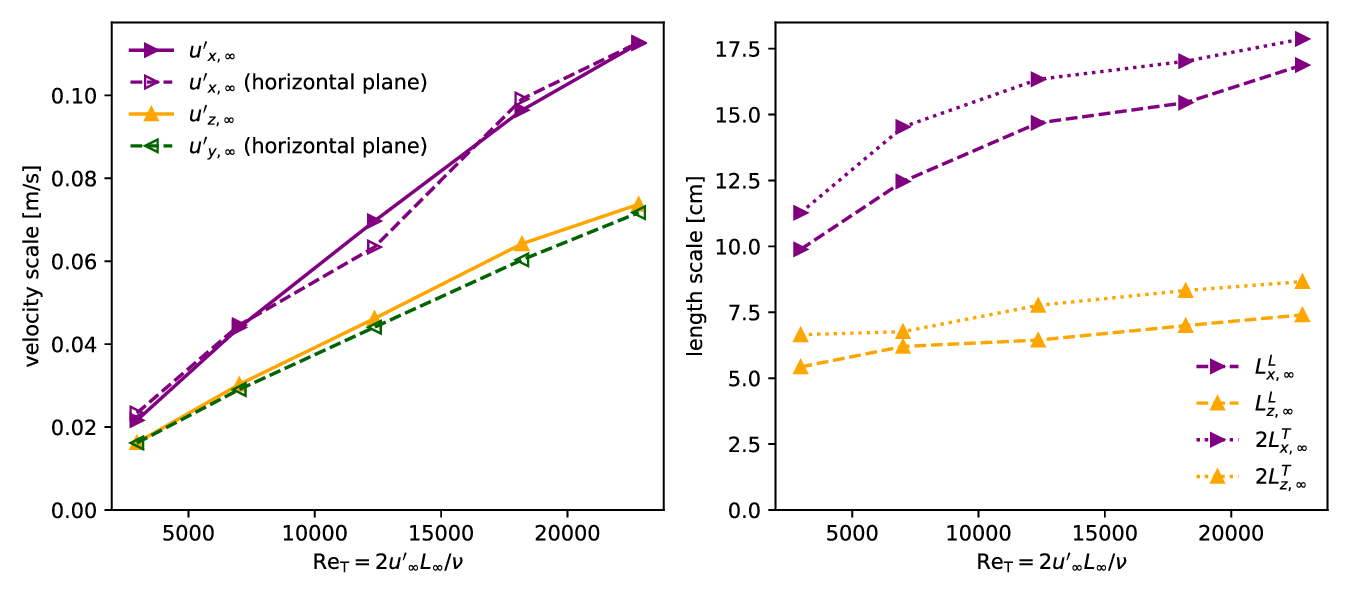}\put(4,43){(a)}\put(52,43){(b)}\end{overpic}}
  \caption{Conditions in the bulk at each Reynolds number: (a) the components of the root mean square velocity fluctuations and (b) the longitudinal and transverse integral length scales.}
\label{fig:bulk_velscale_lenscale}
\end{figure}

In both the horizontal (surface-parallel) direction ($i = x$) and the vertical (surface-normal) direction ($i = z$), the turbulent velocity field is Reynolds-decomposed as $U_i (x,z,t)= \overline{U_i} (x,z) + u_i (x,z,t)$, where $\overline{U_i}$ is the local mean and $u_i$ is the instantaneous fluctuation. Figure \ref{fig:bulk_velscale_lenscale} (a) shows the components of the fluctuating velocity in the bulk, $u'_{i,\infty}$, for each case, displaying a level of large-scale anisotropy typical of similar setups \citep{esteban_laboratory_2019}. 

The four available components of the spatial autocorrelation tensor can be calculated as
\begin{equation}
    \rho_i^s\dann{(r)} = \frac{ \overline{u_i(\vec{x}) u_i(\vec{x} + r \vec{e}_{i,s})} }{u'_i(\vec{x})^2}, \label{eq:autocorrelation_tensor}
\end{equation}
where \new{$s$ signifies whether a longitudinal ($s=\mathrm{L}$) or transverse ($s=\mathrm{T}$) correlation is considered and $i$ signifies the component ($x$ or $z$) of the velocity considered. Thus, $\vec{e}_{i,\mathrm{L}} = \vec{e}_i$ in order to obtain a longitudinal correlation, and $\vec{e}_{i,\mathrm{T}}$ is orthogonal to $\vec{e}_i$ in order to obtain a transverse one (specifically, using $\vec{e}_{x,\mathrm{T}}=-\vec{e}_z$ and $\vec{e}_{z,\mathrm{T}}= \vec{e}_x$).} \new{Horizontal} homogeneity warrants independence from the generic position $\vec{x}$ in the measurement plane \new{and enables replacing $\vec{e}_x$ with $-\vec{e}_x$}.

Figure \ref{fig:bulk_velscale_lenscale} (b) shows the four integral length scales in the bulk, $L_{i,\infty}^s$, at each Reynolds number, found by identifying the $1/e$ crossing of the corresponding \new{component of equation \ref{eq:autocorrelation_tensor}} (or with integration in the case of $L_{z,\infty}^\mathrm{T}$, given its quick convergence). As \new{these} length scales are associated to the width attained by the jets in the homogeneous turbulence region at the centre of the tank, they are weakly sensitive to the power supplied to the pumps \citep{carter_generating_2016}. \new{To more easily assess the isotropy, the} integral scales based on the transverse autocorrelations are shown multiplied by two according to the relation for homogeneous isotropic turbulence, $L_{i,\infty}^\mathrm{L} = 2 L_{i,\infty}^\mathrm{T}$ \citep{pope_turbulent_2000}. The jet-driven forcing causes the horizontal velocity fluctuations to remain correlated over larger distances (both longitudinal and transverse) compared to the vertical fluctuations \citep{carter_scale--scale_2017,esteban_laboratory_2019}.

Additional PIV measurements are performed along a horizontal plane at $z=\SI{-20}{cm}$, using similar hardware and achieving similar resolution as in the near-surface vertical planes. Figure \ref{fig:bulk_velscale_lenscale} (a) shows, \new{with the dashed lines,  the values of $u'_{x,\infty}$ and $u'_{y,\infty}$ calculated from the 2000 snapshots per condition taken during} these measurements; comparison between $u'_{y,\infty}$ and $u'_{z,\infty}$ confirms that velocity statistics in the $y$—direction are quantitatively similar to those in the $z$—direction far from the surface. \new{Given the similarity of the velocity statistics in these two directions}, for some \dan{statistical} vectorial quantity  $p_i$ in the bulk we assume $p_y \approx p_z$  and define a characteristic scalar value as
\begin{equation}
    p = \sqrt{\frac{p_x^2 + 2 p_z^2}{3}}.
\end{equation}
At the present levels of anisotropy, alternative strategies of directional averaging \new{(such as taking an algebraic average or neglecting the anisotropy altogether)} yield marginally different values \citep{carter_generating_2016}.

\begin{figure}
  \centerline{\begin{overpic}[width=\linewidth]{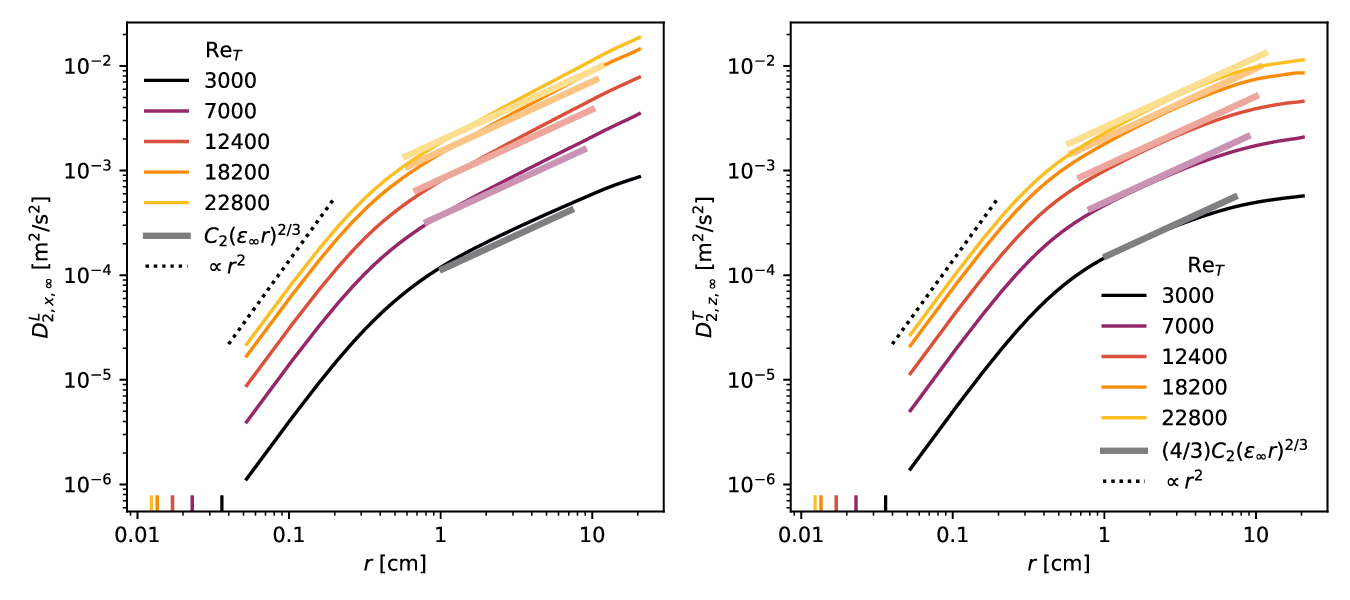}\put(4,43){(a)}\put(52,43){(b)}\end{overpic}}
  \caption{Longitudinal (a) and transverse (b) structure functions for horizontal separations in the bulk for each Reynolds number. Solid thick lines give the inertial range scaling given $\epsilon_\infty$, and dotted lines give the $\propto r^2$ scaling. }
\label{fig:bulk_structure_functions}
\end{figure}

We further compute the $n$-th-order structure function as
\begin{equation}
    D_{n,i}^s (r,z) = \overline{ \left( u_i(\vec{x}+r \vec{e}_{i,s}) - u_i (\vec{x}) \right)^n },
\end{equation}
with $s$ used as in equation \ref{eq:autocorrelation_tensor}. The second-order structure functions based on horizontal separations in the bulk are shown in figure \ref{fig:bulk_structure_functions}, comparing with \citet{kolmogorov_local_1941} predictions in the inertial subrange, $D_{2,x,\infty}^\mathrm{L} = C_2 (\epsilon_\infty r)^{2/3}$ and $D_{2,\new{z},\infty}^\mathrm{T} = (4/3) C_2 (\epsilon_\infty r)^{2/3}$, with $\epsilon_\infty =0.5 {u'_\infty}^3 /L_\infty$ and $C_2=2.0$, which holds for the present range of Reynolds numbers \citep{burattini_normalized_2005,carter_generating_2016,carter_scale--scale_2017,carter_multi-scale_2020}. The Kolmogorov scale in the bulk, $l_\mathrm{K}=(\nu^3⁄\epsilon_\infty )^{1/4}$, is marked in the abscissa of each plot. The curves exhibit the scaling $D_2 \propto r^2$ in the dissipation range, suggesting that the fine scales of the flow are appropriately captured.

Table \ref{tab:kd} summarizes the main properties of the turbulence in the bulk for the considered cases. As confirmed by figure \ref{fig:bulk_structure_functions}, for all cases the Taylor-scale Reynolds number $\mathrm{Re}_\lambda=L_{\lambda,\infty} u'_\infty/\nu$ (with $L_{\lambda,\infty}=u'_\infty \sqrt{15\nu/\epsilon_\infty}$ the Taylor length scale) is sufficiently large to develop an inertial sub-range. The Kolmogorov scales are under-resolved by PIV in the most intense turbulence, but this will not affect the conclusions. For comparison, the bulk turbulence properties of selected previous experimental studies are also listed.

\begin{table}
  \begin{center}
\def~{\hphantom{0}}
\begin{tabular}{lccccccc}
                               & $\ReT$     & $L_\infty$ & $u'_\infty$ [\si{cm/s}] & $\epsilon_\infty$ [\si{m^2/s^3}] & $l_{\mathrm{K}}$ [\si{mm}] & $L_{\lambda}$ [\si{mm}] & $\mathrm{Re}_\lambda$ \\
\multirow{5}{*}{Present study} & 3000       & 7.3        & 1.8                     & \num{4.2e-5}                     & 0.35                              & 10.3                           & 212                   \\
                               & 7000       & 9.0        & 3.6                     & \num{2.5e-4}                     & 0.23                              & 8.2                            & 328                   \\
                               & 12400      & 10.0       & 5.5                     & \num{8.4e-4}                     & 0.17                              & 6.9                            & 432                   \\
                               & 18200      & 10.6       & 7.7                     & \num{2.1e-3}                     & 0.14                              & 6.1                            & 524                   \\
                               & 22800      & 11.6       & 8.9                     & \num{3.0e-3}                     & 0.12                              & 5.9                            & 590                   \\
\citet{brumley_near-surface_1987}                        & 366        & 2.49       & 0.76                    & \num{8.8e-6}                     & 0.53                              & 9.4                            & 80                    \\
\citet{mckenna_role_2004}                        & \makecell{282 \\-- 974}  & 2.5        & \makecell{0.56 \\-- 2.92}            & \makecell{\num{3.5e-6} \\  -- \num{5.0e-4}}      & \makecell{0.19 \\-- 0.67}                      & \makecell{4.8 \\-- 10.9}                    & \makecell{69 \\-- 157}             \\
\citet{herlina_experiments_2008}                        & \makecell{260 \\-- 780} & \makecell{2.8 \\-- 2.9}  & \makecell{0.46 \\-- 1.40}            & \makecell{ \num{1.7e-6} \\ -- \num{4.8e-5}}     & \makecell{0.3 \\-- 0.8}                       & \makecell{7.4 \\-- 12.9}                    & \makecell{66 \\ -- 116}              \\
\citet{variano_turbulent_2013}                        & 6440       & 7.57       & 4.3                     & \num{5.2e-4}                     & 0.19                              & 6.5                            & 314                  
\end{tabular}
  \caption{Properties of the turbulence in the bulk with the forcing conditions employed. Values from selected previous experimental studies are also listed. When not reported, the properties in those studies are deduced from the presented information.}
  \label{tab:kd}
  \end{center}
\end{table}

\begin{figure}
  \centerline{\includegraphics[width=0.75\linewidth]{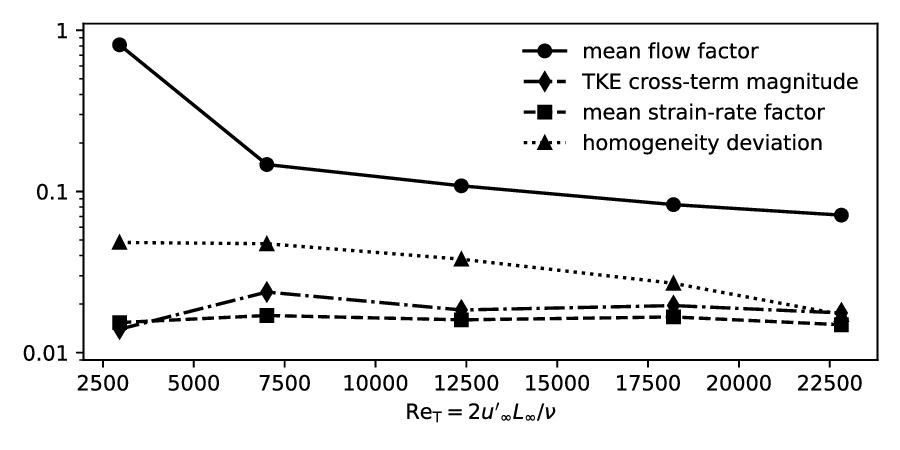}}
  \caption{\new{Parameters related to the homogeneity and isotropy of the bulk flow, defined in eqs. \ref{eq:mean_flow_factor} -- \ref{eq:homogeneity_deviation}.}}
\label{fig:MFF_MSRF}
\end{figure}

Compared to oscillating-grid systems featured in most previous experimental studies of sub-surface turbulence, the present setup produces substantially smaller mean recirculation and inhomogeneities over a larger region \citep{mckenna_role_2004,blum_effects_2010,bellani_homogeneity_2014,carter_generating_2016}. Various metrics to characterize the approximation to zero-mean flow homogeneous turbulence are presented in figure \ref{fig:MFF_MSRF}. In particular, following \citet{carter_generating_2016,esteban_laboratory_2019}, we calculate: the mean flow factor, which is the magnitude of the mean flow relative to the turbulent fluctuations; \new{the relative magnitude of the resolved cross-term in the TKE ($\overline{u_x u_z}$), which is 0 in isotropic turbulence}; and the mean strain-rate factor, which compares the strain rate of the mean flow and the turbulent strain rates. These quantities are defined with
\begin{align}
    \textrm{mean flow factor} &= \frac{\sqrt{\overline{U_x}^2 + 2 \overline{U_z}^2}}{ \sqrt{ {u'_x}^2 + 2 {u'_z}^2 }}, \label{eq:mean_flow_factor}\\
    \textrm{TKE cross-term magnitude} &= \frac{| \overline{u_x u_z} | }{ \overline{u_x^2} + 2\overline{u_z^2}},\label{eq:TKE_cross_term_magnitude}\\
    \textrm{mean strain-rate factor} &= \frac{\sqrt{(\partial \overline{U_x} / \partial x)^2 + 2 (\partial \overline{U_z}  / \partial z)^2}} { \sqrt{ \overline{(\partial u_x / \partial x)^2} + 2\overline{(\partial u_z / \partial z)^2} }},\label{eq:mean_strain_rate_factor}
\end{align}
with all quantities first computed locally and then spatially averaged over the bulk region $z<\SI{-15}{cm}$. The latter quantity is especially important to distinguish the canonical case of homogeneous turbulence from situations in which mean velocity gradients are significant (as in open-channel flows and shallow riverine environments, \citep{nezu_turbulence_1993}). Further, we quantify the homogeneity deviation throughout the bulk region by dividing the standard deviation of the local values of $u'$ by the characteristic $u'_\infty$ (ie, the average of all such local $u'$ values), 
\begin{equation}
    \textrm{homogeneity deviation} = \mathrm{std}(u')_\infty / u'_\infty. \label{eq:homogeneity_deviation}
\end{equation}

We find values of the $\textrm{mean strain-rate factor} < 0.02$, indicating that nearly all the dissipation occurring is turbulent, \new{the $\textrm{homogeneity deviation} < 0.05$, indicating good spatial homogeneity, and the $\textrm{TKE cross-term magnitude} < 0.03$, indicating that we can well-approximate the total TKE while neglecting the turbulence anisotropy}. With the exception of the lowest $\ReT$ case, the mean flow is also relatively weak, $\mathrm{MFF} < 0.2$. It is worth stressing that those qualities, in particular homogeneity, are obtained over a region larger than the integral scale of the turbulence, which is essential for establishing the natural energy cascade \citep{bellani_homogeneity_2014,carter_generating_2016}.

\subsection{Free-surface deformation}

\begin{figure}
  \centerline{\begin{overpic}[width=1\linewidth]{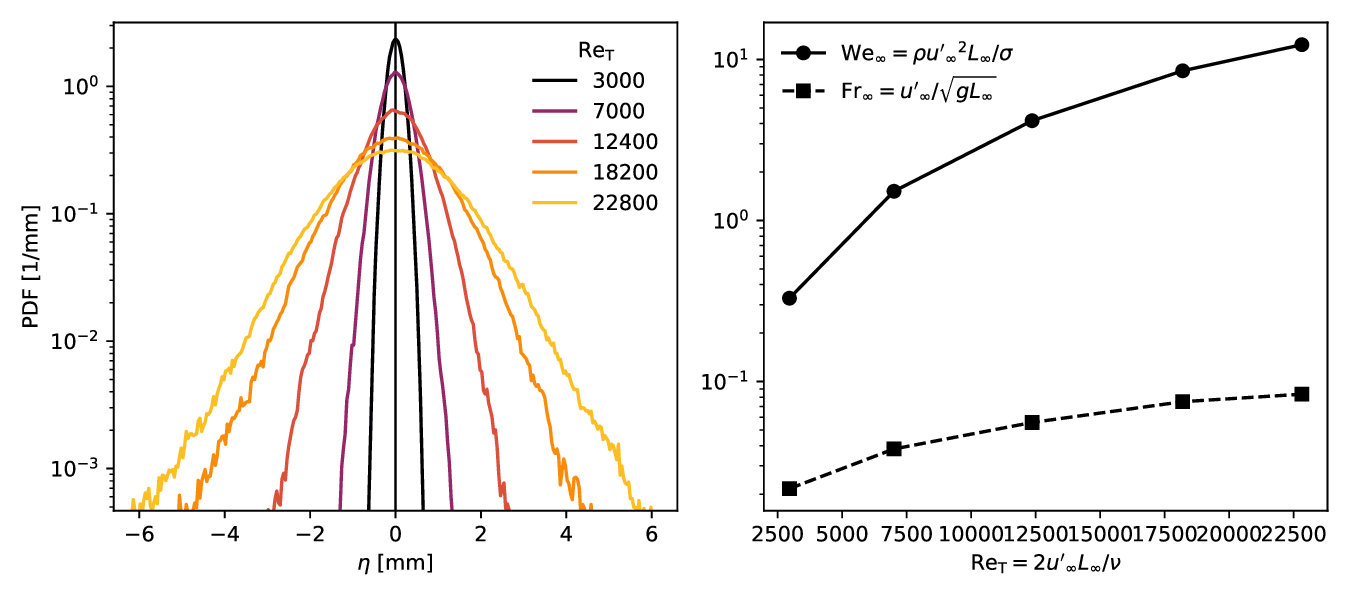}\put(4,43){(a)}\put(52,43){(b)}\end{overpic}}
  \caption{Deformations to the free surface. (a) Distributions of the surface elevation at each condition. (b) Weber and Froude numbers calculated with values from the bulk at each condition.}
\label{fig:surface_deformations}
\end{figure}

Figure \ref{fig:surface_deformations} (a) shows probability density functions (p.d.f.s) of the surface elevation $\eta$ obtained by LIF for each condition. The scale of the surface disturbances, estimated as $2\eta'$, increases with $\ReT$ and is limited to approximately \SI{3}{mm} in the most intense turbulence. Further, figure \ref{fig:surface_deformations} (b) plots the bulk Weber number $\mathrm{We}_\infty = \rho {u'_\infty}^2 L_\infty / \sigma$ and the bulk Froude number $\mathrm{Fr}_\infty =u'_\infty /\sqrt{g L_\infty }$ (with $g$ the gravitational acceleration), which characterize the ability of the large-scale turbulent motions to deform the surface against the restoring action of surface tension and gravity, respectively. Even at the larger $\ReT$, while turbulence is strong enough to counteract surface tension ($\mathrm{We}_\infty > 1$), the large spatial scales guarantee $\mathrm{Fr}_\infty < 1$. In this regime of “gravity-dominated turbulence” \citep{brocchini_dynamics_2001-1}, the surface is expected to display small deformations, coherent with the distributions shown in figure \ref{fig:surface_deformations} (a).

In section \ref{sec:vertical_fluctuations} we show that, below a thin near-surface layer barely resolved by the imaging system, the orbital velocities induced by gravity-capillary waves are small compared to the turbulent velocities we measure. Nonetheless, the surface information obtained from the LIF images is critical in the experimental data processing, as it enables us to mask out the noisy region of the PIV images above the surface.

\section{Turbulence modulation by the free surface}
\label{sec:turbulence_modulation}

\subsection{Vertical fluctuations}
\label{sec:vertical_fluctuations}

\begin{figure}
  \centerline{\begin{overpic}[width=1\linewidth]{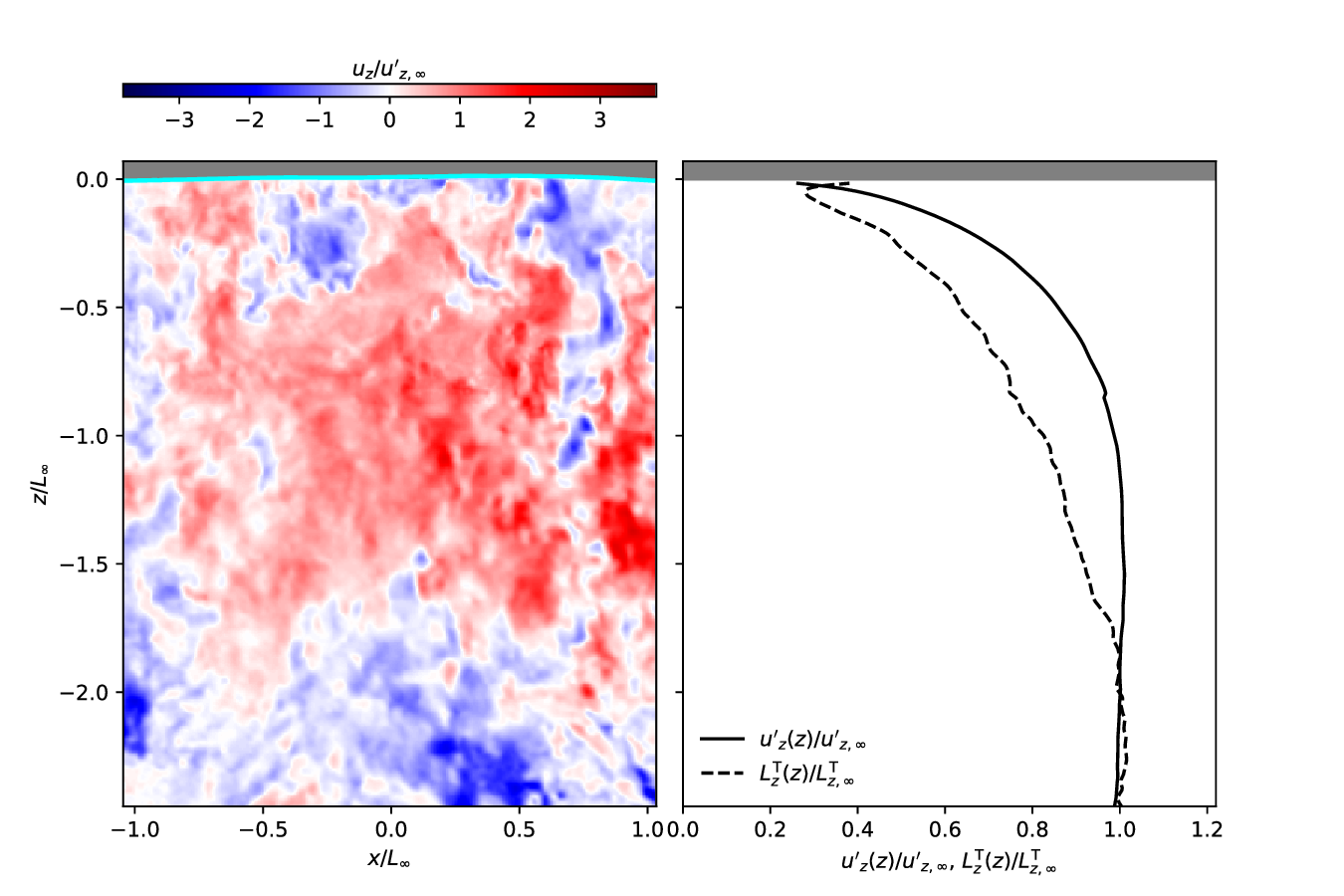}\put(4,56){(a)}\put(52,56){(b)}\end{overpic}}
  \caption{Structure of the vertical velocity fluctuations. (a) Vertical velocity field at one instant in time. (b) Profiles of the vertical fluctuation velocity scale (solid line) and horizontal correlation length of vertical velocity (dashed line) as functions of depth, normalized by their values in the bulk. }
\label{fig:uzfield_and_decays_power50}
\end{figure}

Consistent with previous works, we observe marked changes in the statistics of the turbulence within the blockage layer, $z>-L_\infty$. Figure \ref{fig:uzfield_and_decays_power50} (a) shows a snapshot of the normalized vertical velocity fluctuation field, $u_z/u'_{z,\infty}$, at $\ReT=12400$. Here and in the rest of the paper, when results are shown for only one case, this $\ReT$ will be used as representative unless otherwise specified. Near the surface, the magnitude of vertical fluctuations decays, as does the horizontal length scale of the vertical velocity structures. This is evident in figure \ref{fig:uzfield_and_decays_power50} (b), which shows the vertical profiles of $u'_z(z)/u'_{z,\infty}$ and $ L_z^\mathrm{T}(z) / L_\infty^\mathrm{T} $, both quantities decreasing by an order of magnitude across the source layer. The increase of $L_z^\mathrm{T}$ for $z > -0.01 L_\infty$ signals the presence of the viscous sublayer and possibly the influence of surface deformation, as described below.

\begin{figure}
  \centerline{\begin{overpic}[width=1\linewidth]{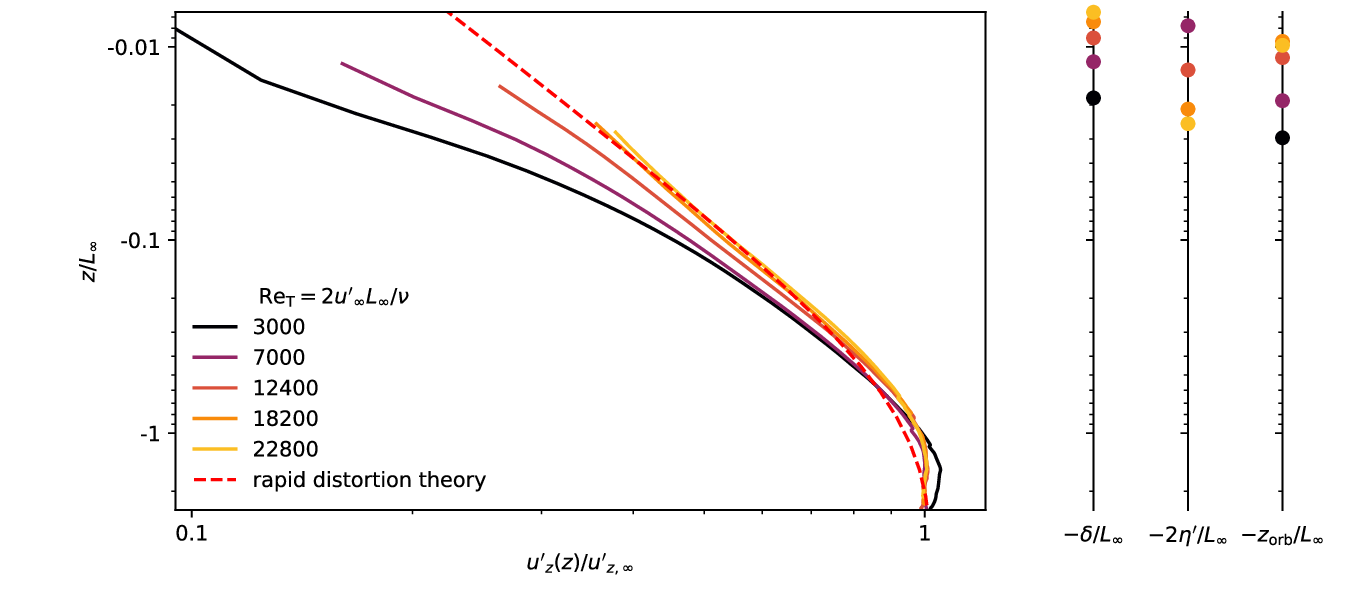}\put(6,44){(a)}\put(77,44){(b)}\put(85,44){(c)}\put(92,44){(d)}\end{overpic}}
  \caption{Decay of the velocity fluctuation scale within the source layer. (a) The vertical velocity fluctuation scale, normalized by its value in the bulk, as a function of depth, normalized by the integral scale in the bulk, for each $\ReT$. The dashed line shows the prediction of rapid distortion theory \citep{hunt_free-stream_1978}. (b-d) The dimensionless positions of the viscous sublayer, intermittency layer, and $-z_\mathrm{orb}$, the depth above which the modeled orbital velocity scale exceeds the measured vertical velocity scale. }
\label{fig:uzprofiles_loglog_allRe}
\end{figure}
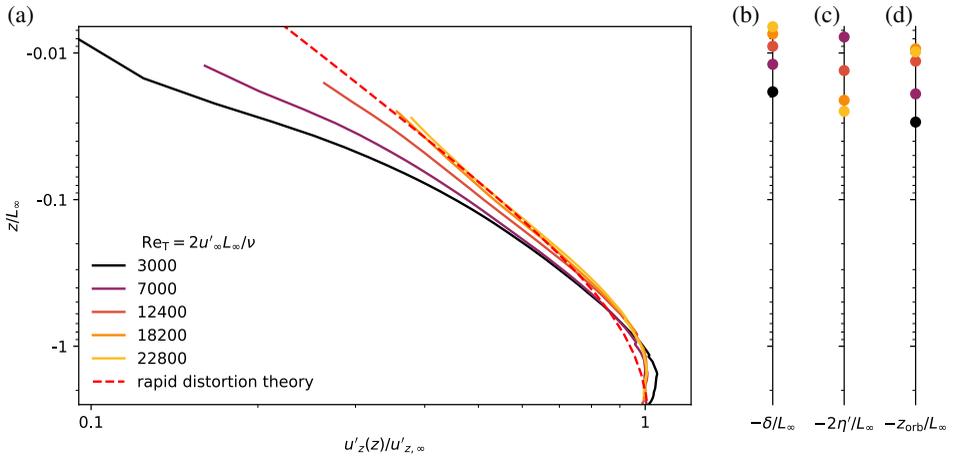

The decay of the vertical velocity fluctuation is shown for all turbulence intensities in figure \ref{fig:uzprofiles_loglog_allRe} (a). With increasing $\ReT$ the trends agree increasingly well with the RDT prediction of \citet{hunt_free-stream_1978}, in particular displaying the scaling $u'_z / u'_{z,\infty} \propto (\dan{-}z/L_\infty)^{1/3}$ . (For this and the following comparisons to their results, we numerically calculate the one-dimensional single-point energy spectra in the source layer according to their equations 2.53-2.55, employing the von Karman spectrum in their eq. 2.63.) This provides strong evidence that the applicability of RDT depends on the turbulence Reynolds number \citep{magnaudet_high-reynolds-number_2003}.

Beside $\ReT$ effects, other factors contribute to the deviation from the power-law relation very close to the surface. First, turbulence statistics change within the viscous sublayer, whose depth is marked in figure \ref{fig:uzprofiles_loglog_allRe} (b) using the estimate 
$\delta_\nu = \ReT^{-1/2} L_\infty $ \citep{brumley_near-surface_1987}. Second, surface deformation results in a so-called “intermittency layer” over which the surface elevation varies in time and space. Following \citet{guo_interaction_2010}, we define this layer as extending to a depth $2 \eta'$ below the mean water level, marked in figure \ref{fig:uzprofiles_loglog_allRe} (c). \new{Third, small} surface undulations \new{generated by the flow} propagate along the surface as capillary-gravity waves \new{(as evidenced by temporally-resolved measurements of $\eta(x,y)$, not reported here, which will be the focus of later work)}, which induce an irrotational orbital velocity $u_\mathrm{orb}$. To gauge the depth $-z_\mathrm{orb}$ over which this is comparable to the turbulent fluctuations, we compute it in a manner inspired by \new{\cite{thais_triple_1995}}. Briefly, each instantaneous surface elevation field is represented by its spatial Fourier transform, and the contribution of each mode to the sub-surface velocity field is computed according to \new{linear wave theory and} the gravity-capillary dispersion relation. We define the depth $-z_\mathrm{orb}$ (shown in figure \ref{fig:uzprofiles_loglog_allRe} (d)) as the height below which $u'_z(z) > \dan{u'}_{\mathrm{orb},z}(z)$. For the representative case $\ReT=12400$, all three types of \new{near-}surface layers have thickness $O(10^{-2} L_\infty)$. In figure \ref{fig:uzprofiles_loglog_allRe} (a) and in the rest of the paper, we display data at $z < -2\eta'$, which does not affect our conclusions.

\begin{figure}
  \centerline{\begin{overpic}[width=1\linewidth,grid=False]{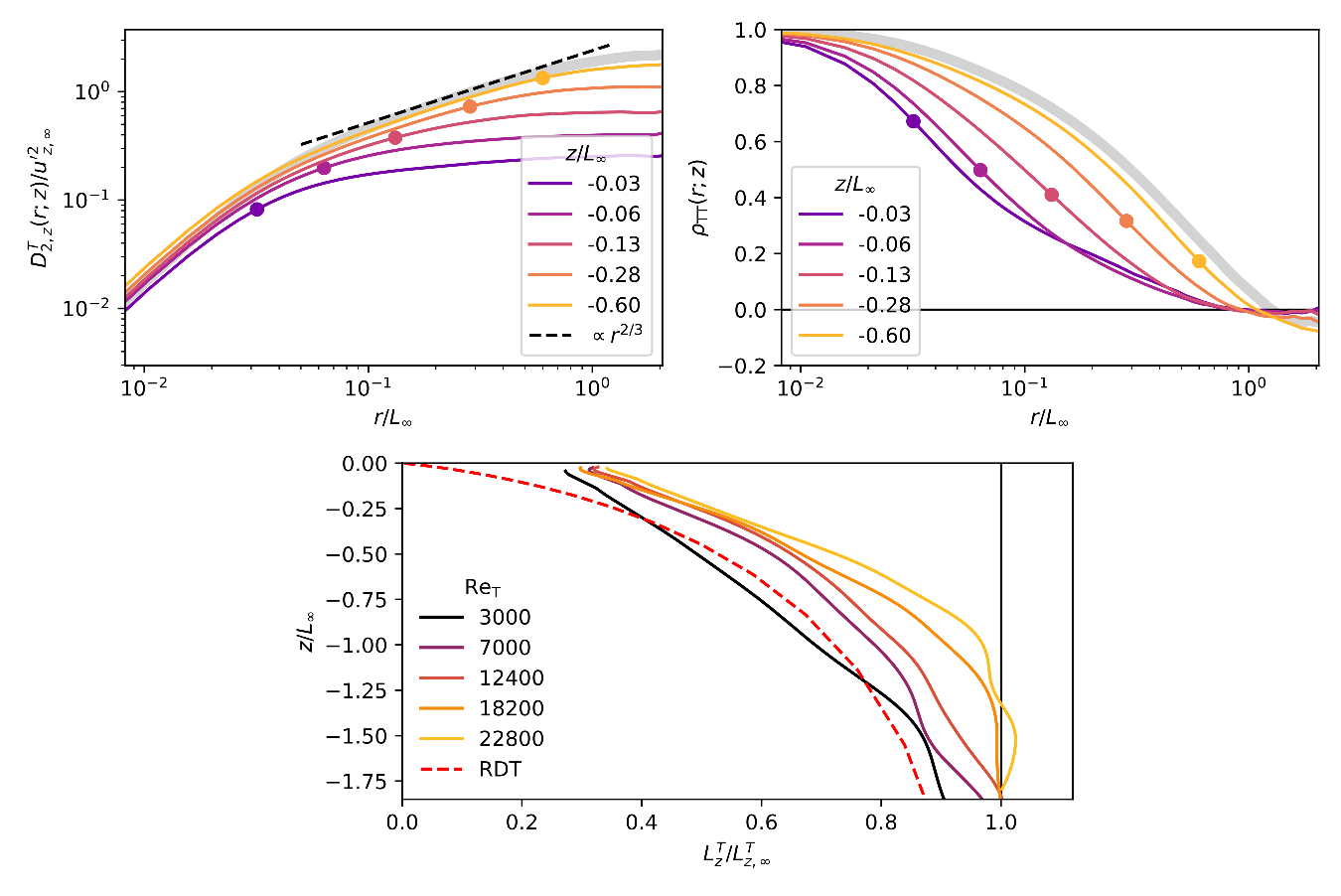}\put(4,64){(a)}\put(51.5,64){(b)}\put(22,32){(c)}\end{overpic}}
  \caption{The spatial structure of vertical motions near the surface. The transverse structure functions (a) and transverse spatial autocorrelations (b) of vertical velocity for various depths with $\ReT = \dan{12400}$. The gray curves give the corresponding values in the bulk. (c) The transverse integral length scale, giving the horizontal footprint of vertical motions, as functions of depth for each Reynolds number. Dashed red curves give rapid distortion theory predictions of \citet{hunt_free-stream_1978}, multiplied by $L_{z,\infty}^\mathrm{T} / (L_\infty/2)$ as a first-order accounting of the anisotropy in the bulk.}
\label{fig:transverse_correlations_vs_z}
\end{figure}

The constraint imposed by the surface on the vertical motions is also manifested in their horizontal structure. This is evident in figure \ref{fig:transverse_correlations_vs_z} (a), in which the transverse structure functions $D_{2,z}^\mathrm{T}$ are plotted at various depths. The circles denote values for $r=-z$, i.e., horizontal separations equal to the depth at which $D_{2,z}^\mathrm{T}$ is calculated. At all depths, the turbulence approximately retains the structure of the bulk at scales $r \leq |-z|$, while the magnitude of the vertical velocity fluctuations is reduced at larger separation. This behaviour is faithfully captured by \citet{hunt_free-stream_1978}'s theory, according to which the transverse spectrum of the vertical velocity component (which carries the same information as $D_{2,z}^\mathrm{T}$) is reduced and flattened at wavenumbers below $|\dan{z}|^{-1}$.

The above suggests that, near the surface, the vertical velocity fluctuations are weakly correlated beyond horizontal scales comparable to the local depth. This is confirmed by figure \ref{fig:transverse_correlations_vs_z} (b), where the data is recast in the form of transverse autocorrelations $\rho_z^\mathrm{T}$. Those decay faster approaching the surface, which corresponds to the decreased $L_z^\mathrm{T}$ shown for all Reynolds numbers in figure \ref{fig:transverse_correlations_vs_z} (c). These compare favourably with the prediction of \citet{hunt_free-stream_1978}, shown as the dashed red line.

\subsection{Horizontal fluctuations}
\label{sec:horizontal_flucs}

Rapid distortion theory predicts an increase in energy in horizontal motions at the expense of that in vertical motions. This has been observed in several experiments on zero-mean-shear flows adjacent to solid boundaries \citep{thomas_grid_1977,johnson_turbulent_2018} and free-surface turbulence simulations \citep{flores_dynamics_2017,guo_interaction_2010,herlina_direct_2014,herlina_simulation_2019} and experiments \citep{brumley_near-surface_1987,variano_turbulent_2013}. It was not observed, however, in the long-time statistics of the decaying turbulence simulations by \citet{perot_shear-free_1995} nor in the experiments by \citet{aronson_shear-free_1997} and by \citet{herlina_experiments_2008}. We hypothesize that the disagreement is due to study-specific characteristics of the bulk turbulence. On one hand, the inviscid RDT analysis assumes a high Reynolds number, which complicated the comparison especially with early simulations. According to \citet{magnaudet_high-reynolds-number_2003}, the relatively low $\ReT$ (resulting in the viscous layer accounting for a significant fraction of the integral scale) was the reason \citet{perot_shear-free_1995} did not observe a near-surface peak of $u_x'\dan{(z)}$ at late times of their decaying turbulence simulations. On the other hand, as mentioned, most experimental studies have applied the forcing to generate the turbulence at distances from the surface much larger than $L_\infty$ (e.g., \citet{mckenna_role_2004,variano_turbulent_2013}). In those systems, any change of turbulent energy approaching the surface is superposed to the spatial decay away from the forcing region. Finally, the ideal conditions of bulk homogeneity, isotropy and zero-mean-shear cannot be fully achieved in experiments, possibly clouding the effect of the surface.

\begin{figure}
  \centerline{\begin{overpic}[width=1\linewidth,grid=False]{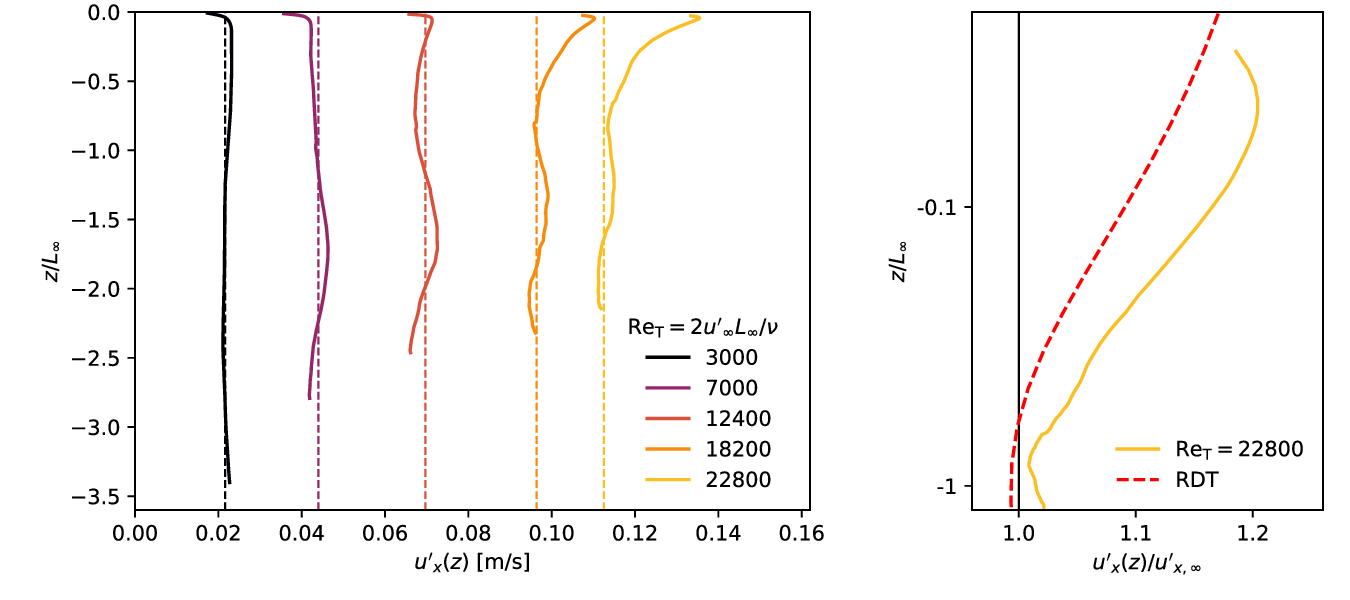}\put(3,44){(a)}\put(68,44){(b)}\end{overpic}}
  \caption{The change in the horizontal velocity fluctuation scale near the surface. (a) Profiles of horizontal r.m.s. velocity fluctuations as functions of depth for each condition, where the dashed line shows the value in the bulk. (b) The same data for the \dan{highest-$\ReT$ case}, non-dimensionalized and compared to the amplification predicted by rapid distortion theory (red, dashed line). }
\label{fig:uxrms_profiles_allRe}
\end{figure}

In the present setup, the distance between the water surface and the axis of the upper-most jets forcing the turbulence is $O(L_\infty)$; thus, the natural spatial decay of energy between the forcing region and the surface is expected to be marginal. Moreover, we are able to assess the influence of the Reynolds number by spanning almost a decade in $\ReT$. Figure \ref{fig:uxrms_profiles_allRe} (a) shows profiles of $u'_x$, indicating how the horizontal energy increase emerges at $\ReT > \sim 10000$, while for weaker forcing it is obscured by spatial inhomogeneities. We remark that this cannot be taken as a general threshold, due to the abovementioned difficulty of comparing different systems. In fact, near-surface amplification of $u'_x$ has been reported in experiments at $\ReT  < 1000$ by \citet{brumley_near-surface_1987}, though with significant scatter.

Figure \ref{fig:uxrms_profiles_allRe} (b) displays the vertical profiles of $u'_x/u'_{x,\infty}$ for the highest Reynolds number, $\ReT = \dan{22800}$, along with the theory of \citet{hunt_free-stream_1978}. The amplification of horizontal energy in our experiments occurs over a greater depth, but the peak is in close agreement with the prediction, $u'_x/u'_{x,\infty} = \sqrt{3/2}$. This is significantly lower than what was observed in numerical studies \citep{walker_shear-free_1996,guo_interaction_2010}, and at least two factors may be responsible. First, in our experiments $u'_x$ decreases in the immediate vicinity of the surface due to the small amount of surfactant (which is practically unavoidable in such configurations \citep{variano_turbulent_2013}). Thus, the peak might be higher in the limit of perfectly clean water. Second, numerical simulations have been conducted at much lower $\ReT$. For reference, \citet{guo_interaction_2010} considered $\ReT =123$, while the most massive simulations to date for this configuration are the ones of \citet{herlina_simulation_2019} at $\ReT = 1856$.

\begin{figure}
  \centerline{\begin{overpic}[width=1\linewidth]{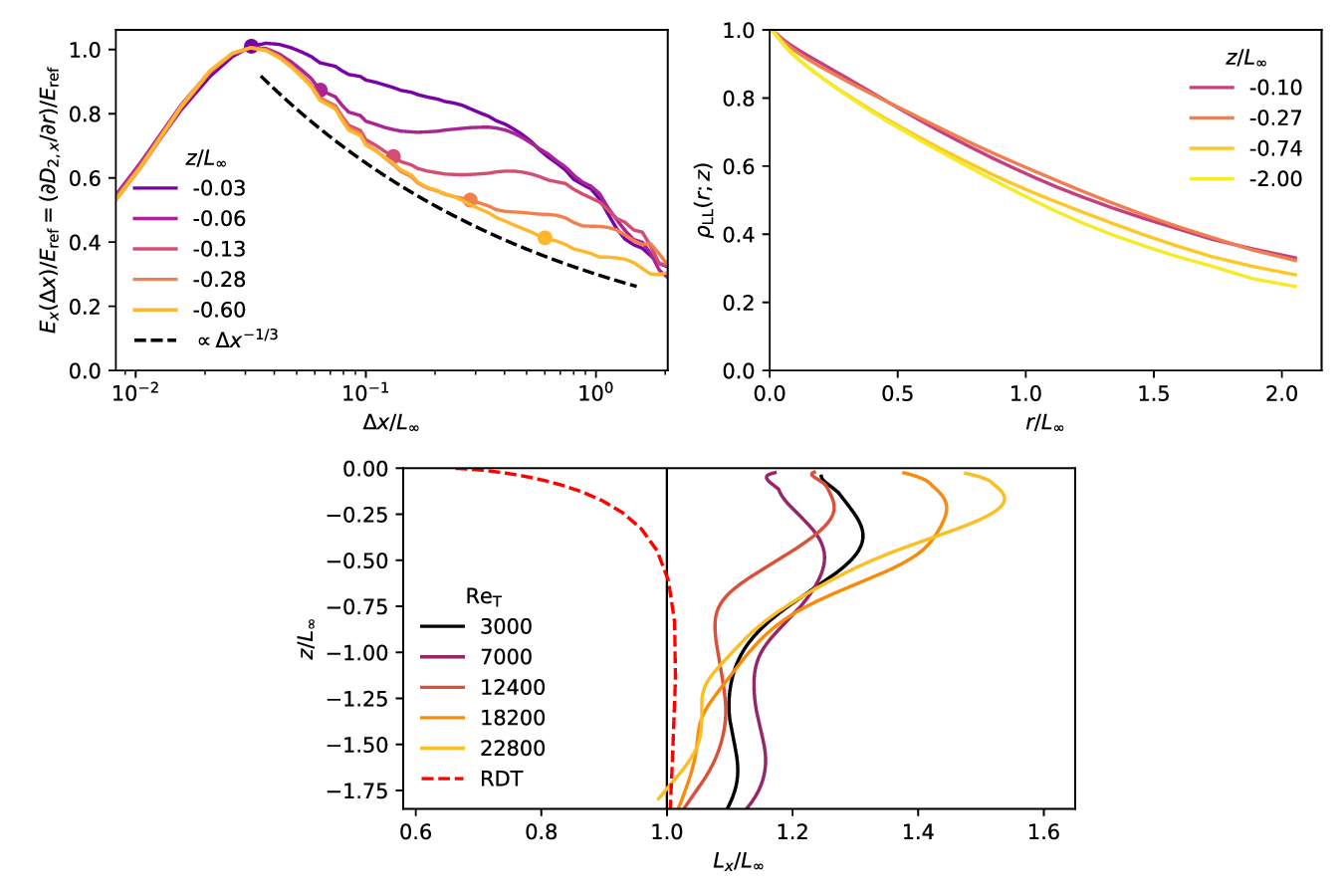}\put(4,64){(a)}\put(50.5,64){(b)}\put(22,32){(c)}\end{overpic}}
  \caption{Structure of surface-parallel velocity. (a) The energy density (the radial derivative of the second-order longitudinal structure function) normalized by its value at $\Delta x = 0.03 L_\infty$. This normalization accounts for the inhomogeneous forcing with depth: such inhomogeneities impact the energy density at all scales, while the surface-induced changes occur solely at scales larger than the local depth. Curves are shown for various $z$, indicating the increased energy density for large scales $\Delta x > -z$. The dashed line shows the Kolmogorov scaling. (b) The longitudinal autocorrelation at various depths. Note that the spatial separations on the horizontal axis are normalized by the characteristic integral scale $L_\infty$ which is shorter than $L_{x,\infty}$ due to the anisotropies in the bulk. (c) The longitudinal integral scale normalized by its bulk value for each Reynolds number as a function of depth. The dashed red line gives the predictions of RDT \citep{hunt_free-stream_1978}, showing that the near-surface amplification we observe experimentally is not captured by this theory.}
\label{fig:longitudinal_correlations_vs_z}
\end{figure}

Having confirmed that, for sufficiently intense turbulence, the horizontal TKE is augmented in the source layer, we explore its scale-to-scale distribution. This is characterized by the horizontal energy density,
\begin{equation}
    \new{E_x (r, z)= \frac{\partial}{\partial r} \left( D_{2,x}^\mathrm{L}(r, z) \right),} \label{eq:horizontal_energy_density}
\end{equation}
which is the scale-space analogue of the energy spectrum \new{at depth $z$ (in that $E_x(r,z) \Delta r$ represents the contributions to the horizontal TKE from structures with size between $r$ and $r+\Delta r$)}. Figure \ref{fig:longitudinal_correlations_vs_z} (a) shows $E_x(r,z)$ at the same depths for which the transverse structure functions are shown in figure \ref{fig:transverse_correlations_vs_z} (a). It is apparent that the spectrum of horizontal energy exceeds the Kolmogorov scaling $E_x \propto r^{-1/3}$ for $r > -z$. Thus, the comparison with figure \ref{fig:transverse_correlations_vs_z} (a) demonstrates how both the augmentation of horizontal energy and the attenuation of vertical energy occur for scales exceeding the local depth. It is notable that the large-scale $E_x (r)$ amplification is evident at all considered $\ReT$---even those for which figure \ref{fig:uxrms_profiles_allRe} (a) shows no appreciable amplification of $u'_x(z)$ near the surface. 

The amplification of horizontal energy at the large scales results in a significant increase in surface-parallel footprint of the near-surface $u_x$ structures. This is demonstrated by the longitudinal autocorrelations $\rho_x^\mathrm{L}$ in figure \ref{fig:longitudinal_correlations_vs_z} (b), which decay more slowly as the surface is approached and result in the evolution of the integral scale $L_x^\mathrm{L} (z)$ in figure \ref{fig:longitudinal_correlations_vs_z} (c). While there is uncertainty due to the limited range over which \new{the employed} exponential fit to the autocorrelations can be performed, there is a substantial increase throughout the source layer, especially at the larger $\ReT$. That is in stark contrast with the theory of \citet{hunt_free-stream_1978}, which predicts a decrease of the correlation length, following $L_x^\mathrm{L} / L_{x,\infty}^\mathrm{L} = ({u'_x}^2 / {u'_{x,\infty}}^2)^{-1}$. \citet{herlina_direct_2014} also observed an increase of $L_x^\mathrm{L}$ approaching the surface, attributing it to the growth of the integral scale as the turbulence decays away from the forcing region \citep{pope_turbulent_2000,davidson_turbulence_2004}. This explanation is less convincing here, as the forcing is applied throughout the sub-surface volume. An alternative explanation is to be found in the way the surface affects the inter-scale transfer of energy, which is discussed in section \ref{sec:inter-scale_energy_transfer}. 

\subsection{Velocity gradients}

\begin{figure}
  \centerline{\begin{overpic}[width=1\linewidth]{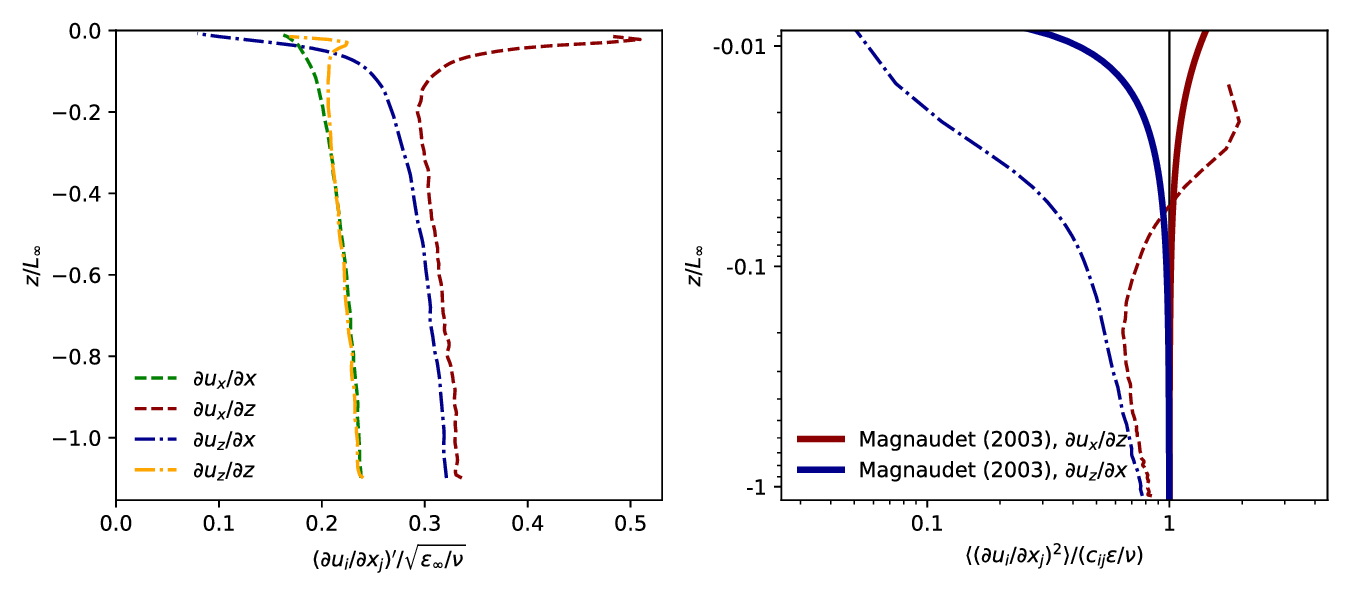}\put(4,43){(a)}\put(52,43){(b)}\end{overpic}}
  \caption{Change in the velocity gradients near the surface for the case with $\ReT = 3000$. (a) The r.m.s. of the four measured components of the velocity gradient tensor. (b) Comparison with the RDT predictions of \cite{magnaudet_high-reynolds-number_2003} for the behavior of the variance of the transverse gradients.}
\label{fig:veloctiy_gradient_profiles_power10}
\end{figure}

The free surface modifies the velocity gradients due to both the kinematic and the dynamic boundary conditions. In figure \ref{fig:veloctiy_gradient_profiles_power10} (a), we plot vertical profiles of the r.m.s. fluctuations for the measured components of the velocity gradients. Here we consider the data for the lowest Reynolds number, $\ReT = 3000$, for which the velocity gradients are best resolved by PIV. The values in the bulk approximately follow the relations for homogeneous isotropic turbulence, $(\partial u_x / \partial \dan{x})' = (\partial u_z / \partial z)' = \sqrt{2} (\partial u_x / \partial z)' = \sqrt{2} (\partial u_z / \partial x)'$ \citep{monin_statistical_1975}. Despite the forcing being applied relatively close to the surface, the r.m.s. velocity gradients still display a weak decay away from the bulk. This is consistent with the fact that small-scale quantities decay faster than large-scale ones, according to established relations for freely decaying turbulence: $\overline{u_i^2} \sim \zeta^{-m}$ and $\overline{(\partial u_i / \partial x_j)^2} \sim \zeta^{-(m+1)}$, where $\zeta$ is the distance from the virtual origin of the forcing and $m = 1 – 1.4$ \citep{hearst_decay_2014,sinhuber_decay_2015}. 

Approaching the surface, $(\partial u_x / \partial x)'$ declines and $(\partial u_z / \partial z)'$ grows. As the viscous layer is approached, they closely approximate the ratio $(\partial u_z / \partial z)'/(\partial u_x / \partial x)' = \sqrt{2}$ predicted by RDT \citep{guo_interaction_2010}. The sharp decrease of $(\partial u_z / \partial x)'$ to negligibly low levels reflects the zero-shear boundary condition, while the increase of $(\partial u_x / \partial z)'$ follows the augmentation of the horizontal fluctuations described above. Overall, the trends are compatible with those reported by \cite{guo_interaction_2010}. However, as the present $\ReT$ is two orders of magnitude larger, the relative thickness of the viscous layer is one order of magnitude smaller, with $\delta_\nu \sim 0.01 L_\infty$ here versus $\delta_\nu \sim 0.1 L_\infty$ in their study. Indeed, the effect of the zero-shear boundary condition (expected to quench $(\partial u_x / \partial z)'$ at the surface) is not reflected by the measurements. Along with imaging limitations, this is due to \new{residual contamination, which permits shear stress at the surface, and surface deformations, which permit enhanced motion along $z=0$.}

In figure \ref{fig:veloctiy_gradient_profiles_power10} (b) our results on the transverse gradients are compared to RDT predictions as obtained by \citet{magnaudet_high-reynolds-number_2003}, which involve an increase in $(\partial u_x / \partial z)'$ and a decrease in $(\partial u_z / \partial x)'$. The measured changes in transverse r.m.s. velocity gradients within the source layer align qualitatively with the theory, though the depth of the affected region and magnitude of the change is underpredicted. The qualitative agreement confirms the significance of the interaction between the fine scales of the turbulence and the large-scale flow modifications imposed by the surface.

\subsection{Inter-scale energy transfer}
\label{sec:inter-scale_energy_transfer}

So far, we have shown a marked change in the density of TKE present at various scales and distances from the free surface, as well as a near-surface change in the dynamics of the small-scale structures. Here we explore how such surface-induced changes are reflected in the transport of energy across scales.

\new{We denote with
\begin{align}
    q_i(\vec{x},\vec{r}) &= (u_i(\vec{x}+\vec{r}) - u_i(\vec{x}))^2,\\
    &= \left(\Delta u_i (\vec{x},\vec{r})\right)^2 
\end{align}
the} TKE associated with the $i$-component of the velocity difference between a point $\vec{x}$ and a point $\vec{r}$ away, \new{noting that $\overline{q_i}(\vec{x},\vec{r}\rightarrow \vec{\infty}) \rightarrow 2 \overline {u_i^2}$ in a homogeneous flow}. With this notation, the rate at which energetic motions are compressed or extended \new{by the local relative motion is $q_i(\vec{x},\vec{r}) \vec{\Delta u} (\vec{x},\vec{r})$}. Averaging in time and along the homogeneous $x$-direction, we obtain with $\overline{q_i \vec{\Delta u}}(z,\vec{r})$ the depth-dependent rate at which \new{TKE due to $i$-direction fluctuations over a separation} $\vec{r}$ is transported \new{between scales}. The approach builds on the generalized Kármán–Howarth equation \citep{von_karman_statistical_1938,monin_statistical_1975,hill_exact_2002}. Even for inhomogeneous and anisotropic flows for which only selected velocity components are captured, this framework provides insight on the magnitude and direction of the energy cascade at the various scales of the turbulence \citep{gomes-fernandes_energy_2015,alves_portela_role_2020,carter_small-scale_2018}.

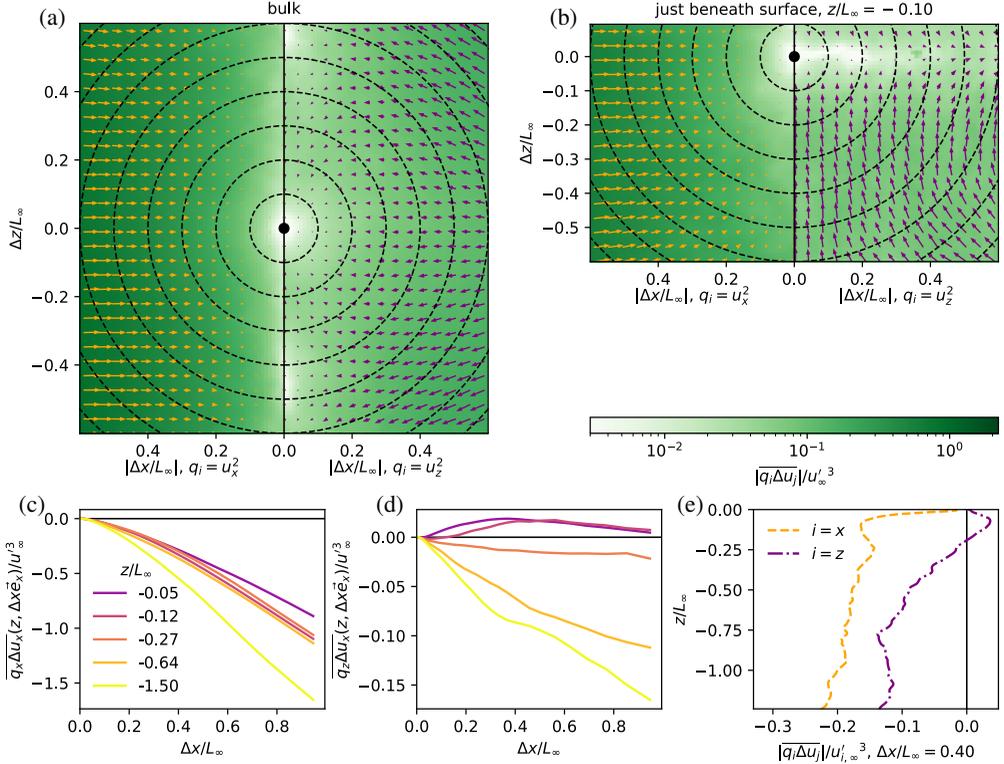
\begin{figure}
  \centerline{\begin{overpic}[width=1\linewidth,grid=False]{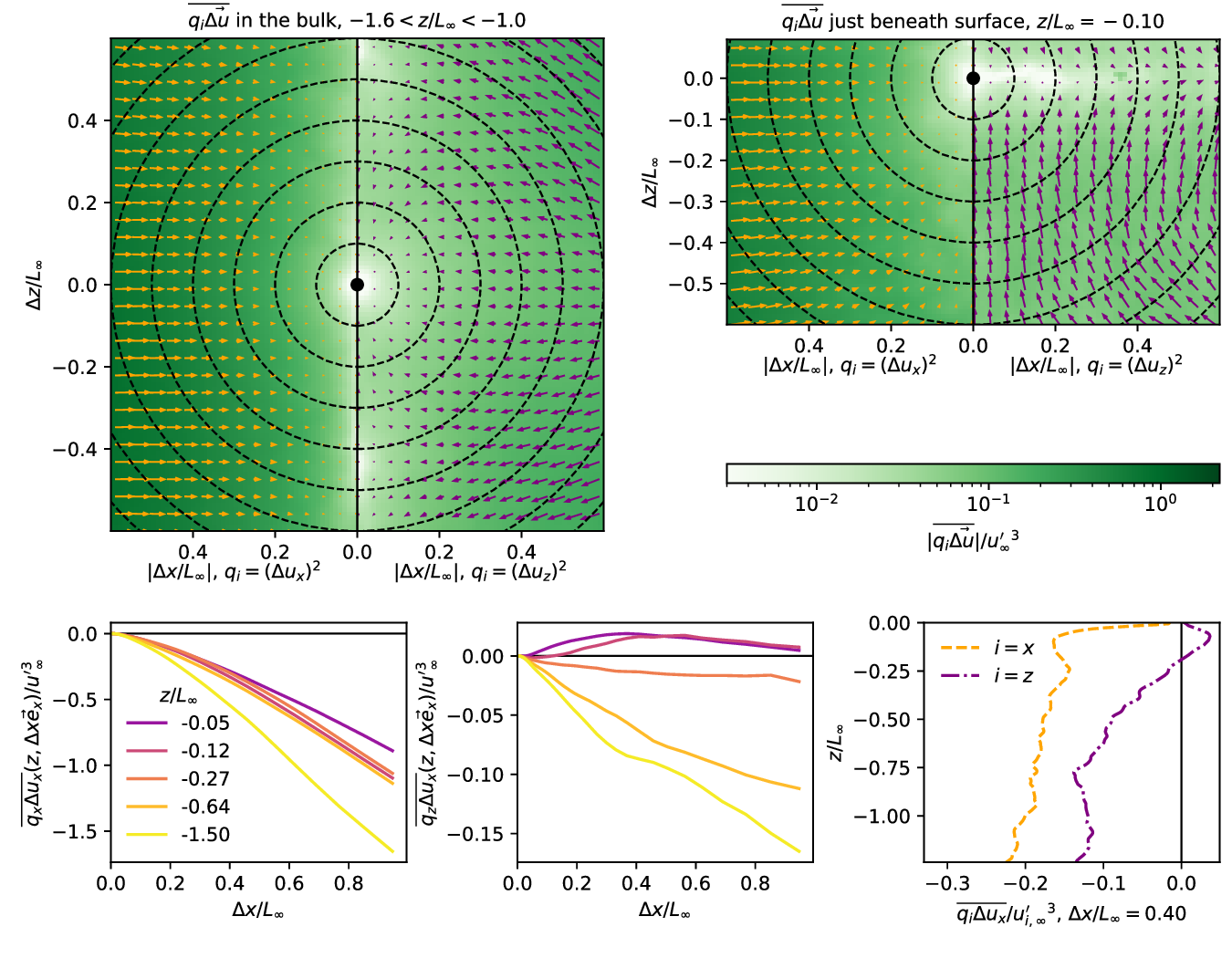}\put(5,74.5){(a)}\put(55,74.5){(b)}\put(3,27){(c)}\put(38,27){(d)}\put(67.5,27){(e)}\end{overpic}}
  \caption{The inter-scale transfer of energy at various depths \new{with $\ReT = 12400$}. (a-b) The vector transfer of energy between scales $\overline{q_i \vec{\Delta u}}(z,\vec{r})$ in the bulk (a) and at $z/L_\infty=-0.1$ (b). In each map, the left and right sides give the inter-scale transfer of horizontal and vertical TKE, respectively. The color gives the magnitude of the transfer and the arrow gives its direction in scale space. Dashed circles trace scales of constant $|\vec{r}| = \sqrt{(\Delta x)^2 + (\Delta z)^2}$. (c-d) The radial component of the horizontal transfer of horizontal (c) and vertical (d) TKE at various depths, indicating a reduction (or even reversal in direction) of the transfer near the surface. \dann{(e) Vertical profiles of the inter-scale transfer of \new{contributions to TKE} for the separation scale $\Delta x / L_\infty = 0.40$.}}
\label{fig:horizontal_energytransfer_unconditional}
\end{figure}

With $\Delta x$ and $\Delta z$ giving the horizontal and vertical components of $\vec{r}$, figure \ref{fig:horizontal_energytransfer_unconditional} (a) shows $\overline{q_i \vec{\Delta u}}(z,\vec{r})$ evaluated in the bulk \new{(averaged between $-1.6 \leq z/L_\infty \leq -1$)}, whereas figure \ref{fig:horizontal_energytransfer_unconditional} (b) shows the same quantity just beneath the surface, evaluated at $z/L_\infty = -0.1$. The left and right part of \dann{both} contour plots refer to the horizontal TKE ($q_x$) and vertical TKE ($q_z$), respectively. The colour and direction of the arrows indicate the magnitude and scale-space direction of transport, respectively. Inwards/outwards-pointing arrows thus indicate compression/extension of the energetic motions, i.e., energy being passed to smaller scales (a direct cascade) or to larger ones (an inverse cascade); see \citet{davidson_turbulence_2004} and \citet{vassilicos_dissipation_2015}.

In the bulk (figure \ref{fig:horizontal_energytransfer_unconditional} (a)), \new{TKE from both horizontal and vertical fluctuations is} primarily transferred inwards---that is, in a direct cascade from larger to smaller scales. \new{However, the the large-scale anisotropy in the flow causes departures from the purely down-scale energy flux \citep{carter_scale--scale_2017}}. The inter-scale transport of $u_x^2$  is greater than that of $u_z^2$, largely because of the greater amount of horizontal TKE available to be transferred, as discussed in detail below. The horizontal compression of both \new{TKE contributions is larger} for the same reason, mirroring results from previous studies focused on flows exhibiting comparable large-scale anisotropy \citep{gomes-fernandes_energy_2015,carter_small-scale_2018}. Further, the anisotropy yields a relatively small energy cascade over vertical separations, consistent with the findings of \citet{carter_small-scale_2018} in a jet-stirred turbulence chamber similar to the present one.

Near the surface (figure \ref{fig:horizontal_energytransfer_unconditional} (b)), the inter-scale transfer is radically different. The magnitude of the horizontal \new{compression} of $u_x^2$ is significantly reduced, and strikingly, the arrows denoting the transfer of $u_z^2$ point outwards for horizontal separations. This indicates that, beneath the surface, there is an inverse cascade of vertical energy: fluid regions \dann{of intense} vertical velocity fluctuations are, in average, stretched horizontally such that \dann{vertical} energy is transferred to larger scales.

These surface-induced modifications to the inter-scale energy transfer are made even more apparent in figure \ref{fig:horizontal_energytransfer_unconditional} (c-e), \new{which show the inwards horizontal} transfer of \new{horizontal and vertical TKE} at various depths. \new{The term giving the stretching of horizontal TKE,} $\overline{q_x \Delta u_x}$, amounts to the longitudinal third-order structure function $D_{3,x}^\mathrm{L}$, whose negative slope is proportional to the inter-scale TKE transfer in the inertial range \citep{davidson_turbulence_2004,vassilicos_dissipation_2015}. As seen in figure \ref{fig:horizontal_energytransfer_unconditional} (c), such negative slope is reduced as the surface is approached, signalling a hindering of the direct cascade. \new{For the stretching of vertical TKE} $\overline{q_z \Delta u_x}$ the trend is even stronger (figure \ref{fig:horizontal_energytransfer_unconditional} (d)): as the surface is approached, the sign of this quantity (and the slope of the curve over an intermediate range of horizontal separations) becomes positive, indicating that the vertical \new{TKE} is, \new{on average,} transferred to larger scales. Figure \ref{fig:horizontal_energytransfer_unconditional} (e) plots \new{vertical} profiles of $\overline{q_i \Delta u_x}(z)$ \new{at the} representative separation $\Delta x/L_\infty = 0.4$ and indicates that, while the general trend \new{of energy transfer reduction} is seen across the source layer, it becomes sharper in the upper stratum of depth $O(0.1 L_\infty)$. Here, the direct cascade of horizontal energy is quenched and the cascade of vertical energy is inverted. 

To understand the origins of this behaviour, it is instructive to Reynolds-decompose both $q_i$ and $\Delta u_x$ and write the horizontal inter-scale transfer as
\begin{equation}
    \overline{q_i \Delta u_x} = q'_i (\Delta u_x)' k_i, \label{eq:stretching_rewritten}
\end{equation}
where $k_i = \overline{q_i \Delta u_x} / (q_i' (\Delta u_x)')$ represents the correlation between $q_i$ and $\Delta u_x$. In other words, the presence of the surface changes the inter-scale energy transfer rate because it affects (i) the turbulent energy available to be transferred $q_i$, (ii) the horizontal extension/compression $\Delta u_x$, and (iii) the correlation between both quantities $k_i$. (Note that $k_x$ contains the same information as the skewness of $\Delta u_x$.) To isolate the effect of the free surface, we express each factor in equation \ref{eq:stretching_rewritten} as the sum of its bulk value and a depth-dependent deviation (denoted with a tilde):
\begin{equation}
    \overline{q_i \Delta u_x} = \left( q'_{i,\infty} + \widetilde{q'_i}(z) \right) \left( (\Delta u_x)'_\infty + \widetilde{(\Delta u_x)'}(z) \right) \left( k_{i,\infty} + \widetilde{k_i}(z) \right),
\end{equation}
where the first, second and third terms on the r.h.s. quantify the effect of the changes in (i), (ii), and (iii), respectively, with depth.

\begin{figure}
  \centerline{\begin{overpic}[width=1\linewidth,grid=False]{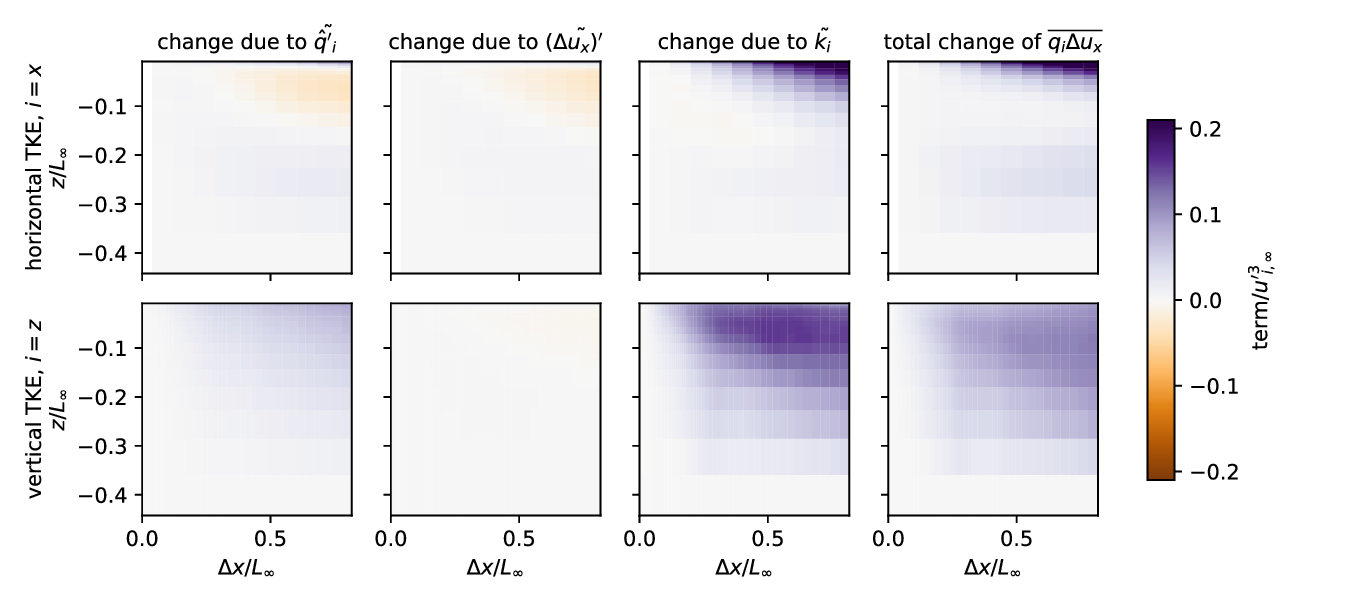}\put(11,38){(a)}\put(29.5,38){(b)}\put(48,38){(c)}\put(66.5,38){(d)}\put(11,20){(e)}\put(29.5,20){(f)}\put(48,20){(g)}\put(66.5,20){(h)}\end{overpic}}
  \caption{Terms relating to the modification of inter-scale energy transfer near the surface for horizontal (a-d) and vertical (e-h) TKE for \dan{$\ReT = 12400$}. Orange values indicate an enhancement of the cascade (i.e., making the transfer more negative), while purple values indicate a reduction of the cascade. Panels (a) and (e) show the effect of the change in energy available to be transferred; panels (b) and (f) indicate the effect of the change in the scale of the velocity differences; panels (c) and (g) indicate the change due to the decorrelation between the energetic structures and horizontal compression. Panels (d) and (h) show the total change in transfer relative to the bulk value. Given that the most pronounced changes occur in the upper half of the source layer and that our results are influenced by forcing-induced anisotropies in the bulk, we take the “bulk” to be $z/L_\infty = -0.4$ for the purpose of this analysis.}
\label{fig:contributions_to_interscale_hinderance}
\end{figure}
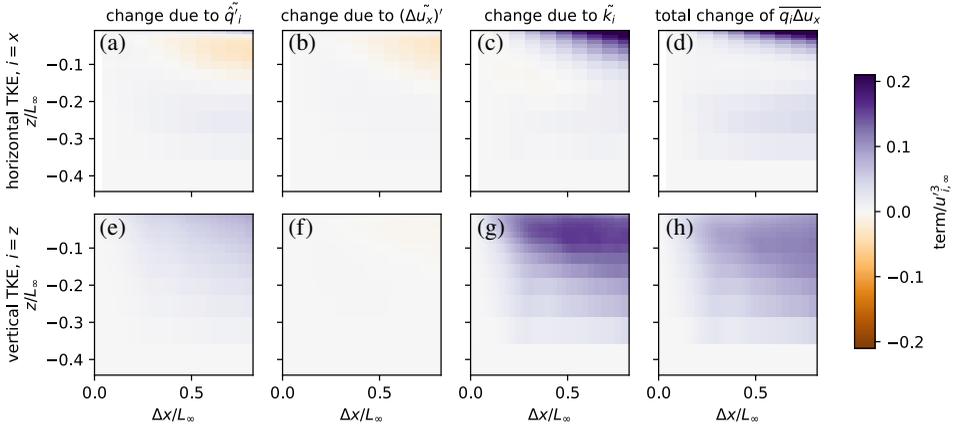

These three contributions and their combined effect on the inter-scale transfer are depicted in figure \ref{fig:contributions_to_interscale_hinderance} as a function of $\Delta x$ and $\Delta z$, for the horizontal transfer of $q_x$ (panels a-d) and $q_z$ (e-h). \new{As discussed in section \ref{sec:vertical_fluctuations}, the horizontal TKE is increased near the surface, especially at large scales.  This increase in the amount of TKE available to be transferred through scales causes the down-scale transfer of $q_x$ to become more negative; see figure \ref{fig:contributions_to_interscale_hinderance} (a). Figure  \ref{fig:contributions_to_interscale_hinderance} (e) and section \ref{sec:horizontal_flucs} show the opposite is the case for $q_z$.} This \new{change in the magnitude of the TKE present}, however, is sub-dominant compared to the decreased coupling between TKE and horizontal extension/compression (figure \ref{fig:contributions_to_interscale_hinderance} (c,f)) which effectively determines the behaviour of the inter-scale transfer for both components (figure \ref{fig:contributions_to_interscale_hinderance} (d,h)). In both cases, the decreased coupling makes $\overline{q_i \Delta u_x}$ less negative, hindering the cascade. The surface also induces a somewhat larger magnitude of $\Delta u_x$ (figure \ref{fig:contributions_to_interscale_hinderance} (b,e)), though this effect is moderate.

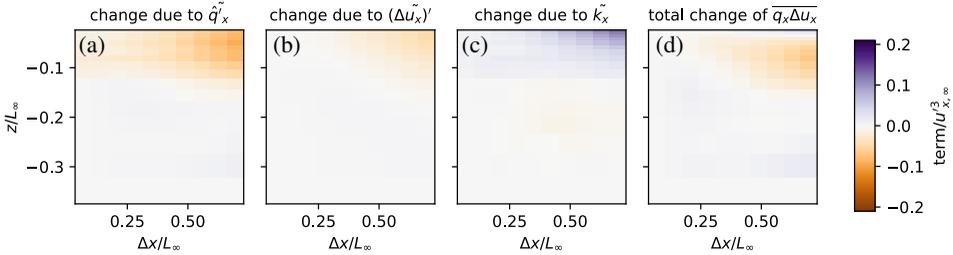
\begin{figure}
  \centerline{\begin{overpic}[width=1\linewidth]{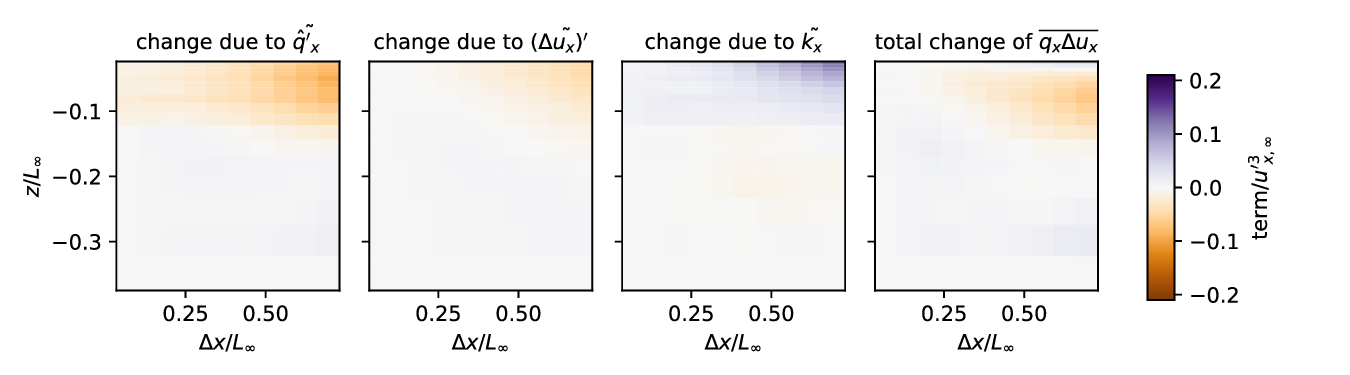}\put(9,21){(a)}\put(28,21){(b)}\put(46.5,21){(c)}\put(65.5,21){(d)}\end{overpic}}
  \caption{As in figure \ref{fig:contributions_to_interscale_hinderance}, but with $\ReT=22800$ and showing only the change in the transfer of horizontal TKE. In this case, the increase in $q_z$ near the surface overcomes the decreased $k_x$, such that the total down-scale energy transfer (d) is enhanced.}
\label{fig:contributions_to_interscale_hinderance_justhTKE_power100}
\end{figure}

So far, we have illustrated the inter-scale energy transfer for the case $\ReT = 12400$. The dynamics of the \new{near-surface cascade of $u_x^2$}, however, \dan{are} sensitive to the degree to which the horizontal TKE accumulates near the surface, which becomes more pronounced with more intense forcing (see figure \ref{fig:uxrms_profiles_allRe}). Figure \ref{fig:contributions_to_interscale_hinderance_justhTKE_power100} illustrates how, for $\ReT = 22800$, the increase in the horizontal energy available to be transferred overcomes the decreased correlation between energetic events and compression. As a result, the net down-scale transfer of horizontal TKE is enhanced near the surface. Still, the behaviour of the vertical TKE transfer (not shown) is qualitatively similar to what is displayed under less intense forcings, with a sizeable backscatter of energy to large scales.

Taken together, the results of this section demonstrate how, along horizontal separations, the surface hinders the direct cascade of horizontal TKE and causes an inverse cascade of vertical TKE. (Isolating the surface-induced changes to the inter-scale energy transfer along vertical separations is more challenging, as those are overwhelmed by the spatial non-homogeneity in this direction.) The hindrance of the direct cascade, \new{we have shown,} stems from the decorrelation between compressive velocity structures and energetic events near the surface. This limits the rate at which large-scale energetic structures break down into smaller eddies and results in the increase of $L_x^\mathrm{L}$ approaching the surface seen in figure \ref{fig:longitudinal_correlations_vs_z} (c). That the same effect is not observed for $L_x^\mathrm{T}$ is likely due to the kinematic boundary condition: this imposes that the horizontal footprint of the vertical fluctuations must approach the one of the surface divergence, whose extent is discussed in the following section.

\section{Role of upwellings and downwellings}
\label{sec:upwellings_downwellings}

We turn to the dynamics of upwellings and downwellings, critical to the transfer of mass and energy between the surface and the bulk \citep{perot_shear-free_1995}. Here we address questions about their magnitude and spatial extent, quantities that are connected to the surface divergence and in turn to the various processes to which the latter is relevant \citep{mckenna_role_2004,magnaudet_turbulent_2006,turney_airwater_2013,kermani_surface_2009,herlina_direct_2014}. We then analyse the role of upwellings and downwellings in the intercomponent energy transfer near the surface, illustrating how the imbalance between both types of events contributes to the inter-scale energy transfer discussed in section \ref{sec:inter-scale_energy_transfer}.

\subsection{Topology and magnitude}

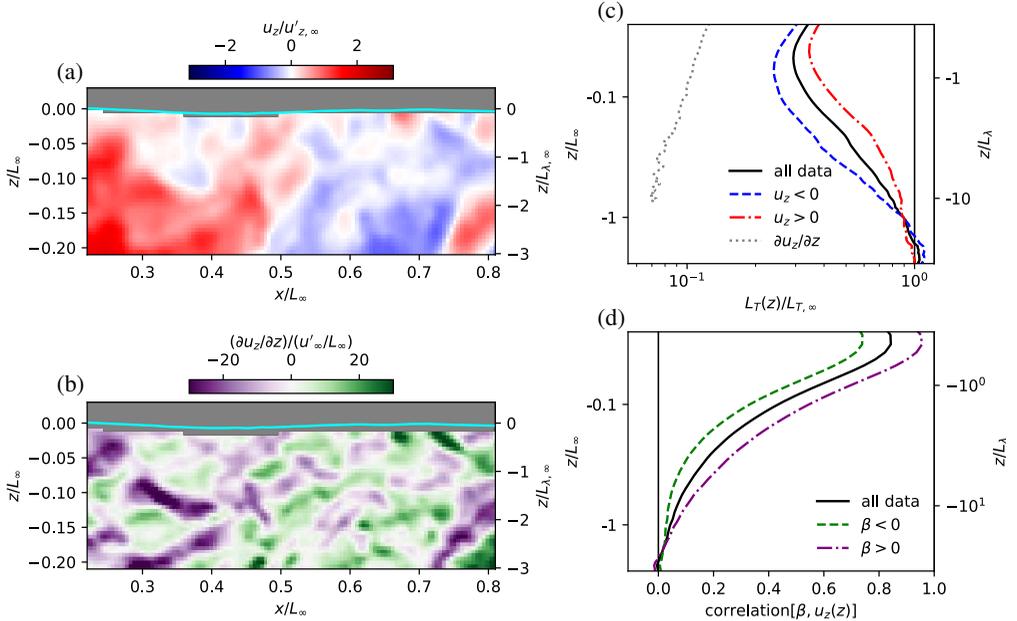
\begin{figure}
  \centerline{\begin{overpic}[width=1\linewidth,grid=False]{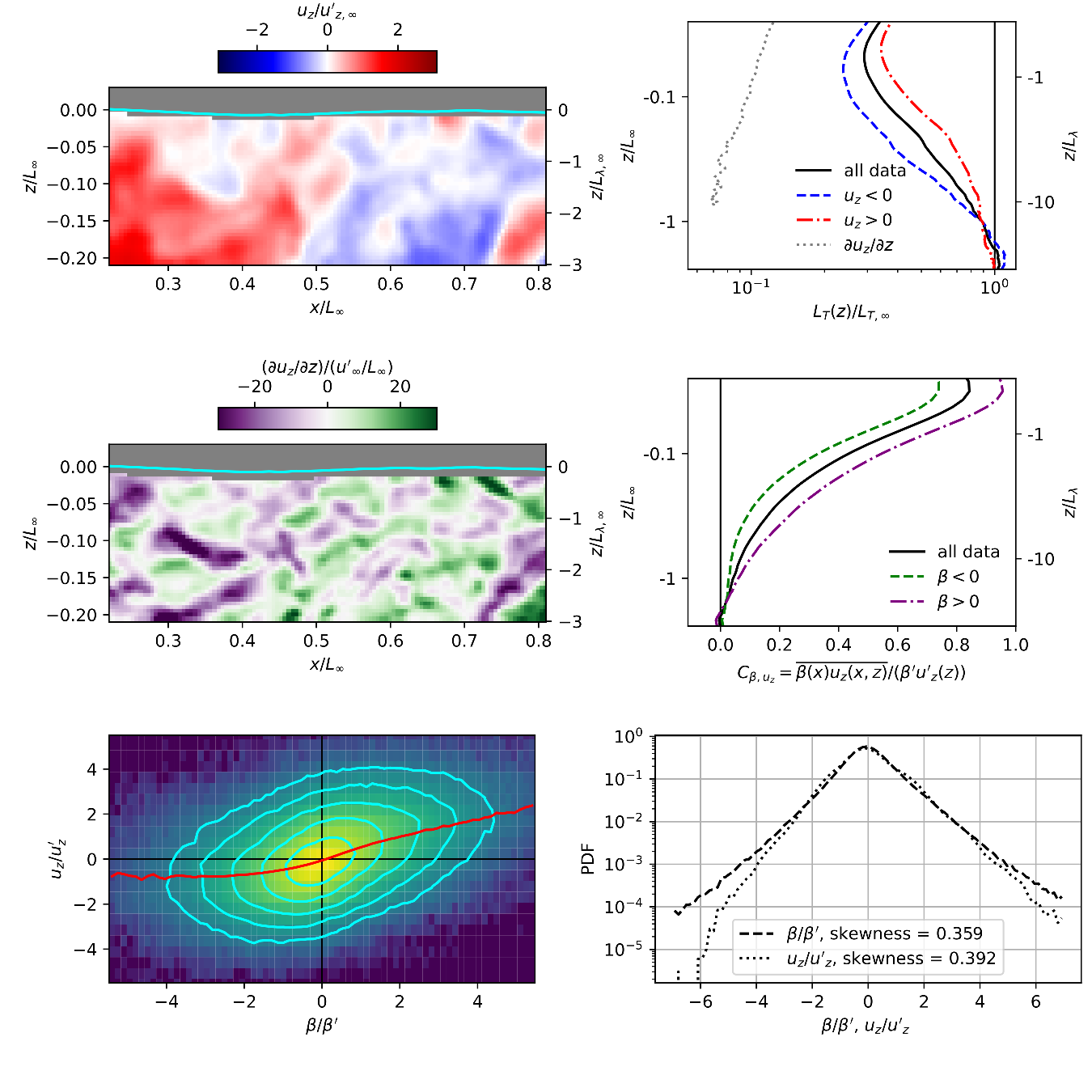}\put(7,93){(a)}\put(60,96){(b)}\put(7,60){(c)}\put(60,63){(d)}\put(7,31){(e)}\put(53,32){(f)}\end{overpic}}
  \caption{The coupling between vertical velocity and its vertical gradient near the surface. (a,c) Snapshots of $u_z$ (a) and $\partial u_z / \partial z$ (c) in the vicinity of the surface. (b) The transverse length scale of the vertical velocity, in black. The blue and red curves give the contributions from downwards and upwards velocities, respectively. The dotted line gray is the horizontal integral scale of $\partial u_z / \partial z$ structures. (d) The correlation between the surface divergence and the vertical velocity as a function of the depth, in black. The purple and green curves give the contributions to the correlation from instances of negative and positive surface divergence, respectively. \new{As in (c), vertical axes are scaled logarithmically.} (e) The joint distribution of surface divergence and sub-surface vertical velocity taken at a depth $z=-L_\infty/10$. Cyan lines trace constant relative occurrences and the red line gives the expected vertical velocity given a surface divergence. (f) Standalone distributions of the two quantities, each normalized by its own standard deviation, evidencing their positive skewness. }
\label{fig:beta_uz_coupling_big}
\end{figure}

The no-penetration condition at the free surface implies that, for small $z$ and an approximately flat surface, $u_z \sim z \partial u_z / \partial z$ \citep{mckenna_role_2004}. This amounts to a positive correlation between the surface divergence $\beta = \partial u_x / \partial x + \partial u_y / \partial y = - \partial u_z / \partial z $ (with the gradients evaluated at the surface) and the sub-surface vertical velocity, making $\beta$ a natural metric to gauge the local state of upwelling/downwelling \citep{guo_interaction_2010}. Figure \ref{fig:beta_uz_coupling_big} (a) and (c) show instantaneous fields of $u_z$ and $\partial u_z / \partial z$, respectively, with the depth normalized by both $L_\infty$ and $L_\lambda$. Within a distance $O(L_\lambda)$ of the surface, there is a \new{resemblance between the two} fields. At larger depths the coherence is gradually lost. 


These visual observations are supported by the flow statistics. Figure \ref{fig:beta_uz_coupling_big} (b) shows how $L_z^\mathrm{T}$ (the characteristic scale of $u_z$ in the homogeneous $x$ direction, \new{shown in black}) shrinks as the surface is approached, while the characteristic scale of $\partial u_z / \partial z$ (based on the integral of its autocorrelation which is $O(L_\lambda)$), \new{grows sightly}. 

The coupling of the surface divergence to the vertical velocity at various depths is quantified \new{for the representative case $\ReT=12400$} in figure \ref{fig:beta_uz_coupling_big} (d), which displays \new{in black} the correlation coefficient between $\beta$ and $u_z$,
\begin{equation}
    C_{\beta, u_z}(z) = \frac{\overline{\beta(x) u_z(x,z)}}{ \beta' u'_z(z)}. \label{eq:C_beta_uz}
\end{equation}
The surface divergence is approximated as \new{$\beta = - \partial u_z/ \partial z$} evaluated at the centre of the uppermost PIV interrogation window, $\sim \SI{1}{mm}$ from the surface. As this is of the order of the viscous sublayer, we expect the estimate to be appropriate to retrieve correct trends \citep{guo_interaction_2010}. The profiles of $C_{\beta, u_z}$ confirm a strong surface-depth correlation near the surface. 

To compare the behaviour during upwellings and downwellings, we condition the statistics on the sign of $\beta$, which is positive for the former and negative for the latter. \new{Given the roughly equal occurrences of $\beta<0$ and $\beta>0$, the overall (unconditioned) correlation between $\beta$ and $u_z$ \new{(equation \ref{eq:C_beta_uz})} could be approximated as the average of the two conditional correlations shown, with
\begin{align}
     \frac{1}{2} \left( \frac{\overline{\beta(\vec{x}) u_z(\vec{x} + z \vec{e}_z) }^-}{\beta' u'_z(z)} + \frac{\overline{\beta(\vec{x}) u_z(\vec{x} + z \vec{e}_z) }^+}{\beta' u'_z(z)} \right) \approx C_{\beta,u_z}(z), \label{eq:surface_depth_correlations_implicit_def}
\end{align}
where the superscripts indicate the sign of $\beta$ on which the averaging is conditioned.} \new{The colored curves in figure \ref{fig:beta_uz_coupling_big} (d), each corresponding to one term in the left-hand side of equation \ref{eq:surface_depth_correlations_implicit_def},} indicate that the surface-parallel flow is correlated to the vertical motion beneath over a deeper depth during upwellings than during downwellings. Likewise, conditioning the transverse \new{covariance} of $u_z$ on its sign yields larger values of $L_z^\mathrm{T}$ when $u_z > 0$ compared to instances when $u_z < 0$; see figure \ref{fig:beta_uz_coupling_big} (b). These results indicate that upwellings have a larger horizontal and vertical extent than downwellings.

The strong correlation between $\beta$ and $u_z$  in the vicinity of the surface is not hindered by the broad distribution of either quantity. This is highlighted in their joint p.d.f. shown in figure \ref{fig:beta_uz_coupling_big} (e), with $u_z$ taken at  $z=-0.1 L_\infty$\dan{,} where $C_{\beta, u_z} \sim 0.5$ (see figure \ref{fig:beta_uz_coupling_big} (d)). The trend of $ \langle u_z \rangle$ conditioned on $\beta$ indicates that upwards sub-surface velocities are more strongly tied to positive divergence than downward ones are to negative divergence. This is true particularly for anomalously large fluctuations. As shown in figure \ref{fig:beta_uz_coupling_big} (f), both $\beta$ and $u_z$ are intermittent and positively skewed: near the surface, fast upwards velocities (thus strongly positive surface divergence) are more likely to occur than fast downwards velocities (and strongly negative surface divergences).

\begin{figure}
  \centerline{\includegraphics[width=1\linewidth]{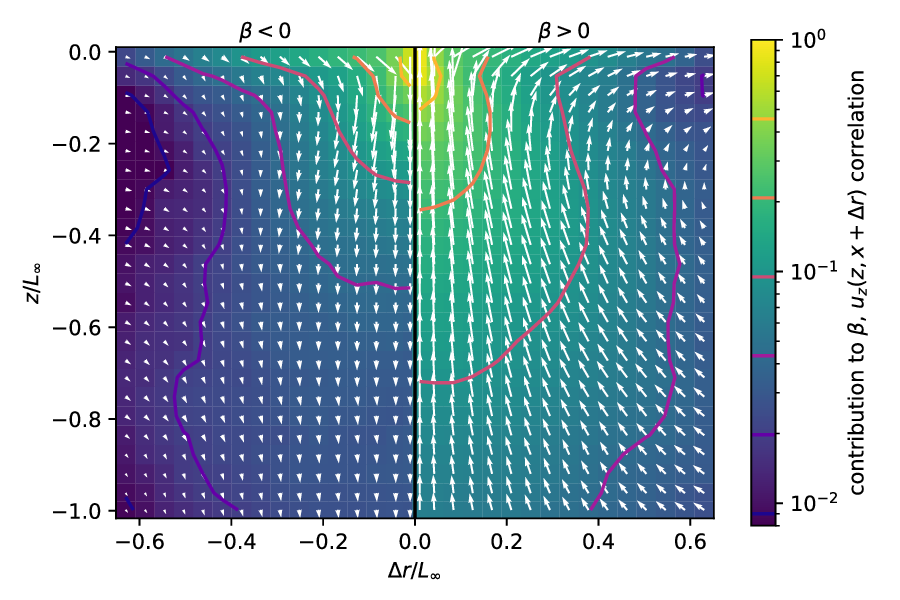}}
  \caption{Contribution to the correlation between $\beta$ and $u_x$ by instances of $\beta < 0$ (left) and $\beta > 0$ (right). The white arrows indicate the weighted-averaged velocity field under each structure.}
\label{fig:beta_uz_spatialcorr_bywelling}
\end{figure}

To complete the view of the flow topology, figure \ref{fig:beta_uz_spatialcorr_bywelling} shows the correlation between $\beta$ and $u_z$ \new{at a depth $z$ and offset horizontally by $\Delta r$}. We condition again on the sign of $\beta$, with contributions from downwellings and upwellings shown on the left and right, respectively. \new{As with the vertical velocity--overhead $\beta$ correlations shown in equation \ref{eq:surface_depth_correlations_implicit_def}, the overall correlation between $\beta$ and the vertical velocity at some depth and horizontal offset can be approximated as
\begin{align}
    \frac{1}{2} \left( \frac{\overline{\beta(\vec{x}) u_z(\vec{x}+\Delta r \vec{e}_x + z \vec{e}_z) }^-}{\beta' u'_z(z)} + \frac{\overline{\beta(\vec{x}) u_z(\vec{x}+\Delta r \vec{e}_x + z \vec{e}_z) }^+}{\beta' u'_z(z)} \right) \approx \frac{\overline{\beta(\vec{x}) u_z(\vec{x}+\Delta r \vec{e}_x + z \vec{e}_z) }}{\beta' u'_z(z)}, 
\end{align}
where the superscripts indicate the sign of $\beta$ on which the averaging is conditioned.} Further, we show with the arrows the \new{mean} sub-surface velocity field conditioned on positive/negative surface divergence, obtained by a conditional weighted average of the sub-surface velocity with $|\beta|$ as the weight. This approach, akin to the variable-intensity spatial averaging schemes employed by \citet{guo_interaction_2010} and \citet{khakpour_transport_2011}, \new{suggests} that \new{coherent} upwellings possess higher intensity and greater spatial extent in both vertical and lateral directions. We stress that this procedure yields a statistical representation of the transport dynamics which is not necessarily representative of instantaneous events--in particular, the averaging smooths the small-scale features of the near-surface fields, such as those pictured in figure \ref{fig:beta_uz_coupling_big} (a,c). \new{Further, with any interpretation of results relating to the size and energetics of upwellings and downwellings, one must keep in mind that, over the entire flow, there is 0 net upwards or downwards mass flux through a given surface-parallel plane}. 

\begin{figure}
  \centerline{\begin{overpic}[width=1\linewidth,grid=False]{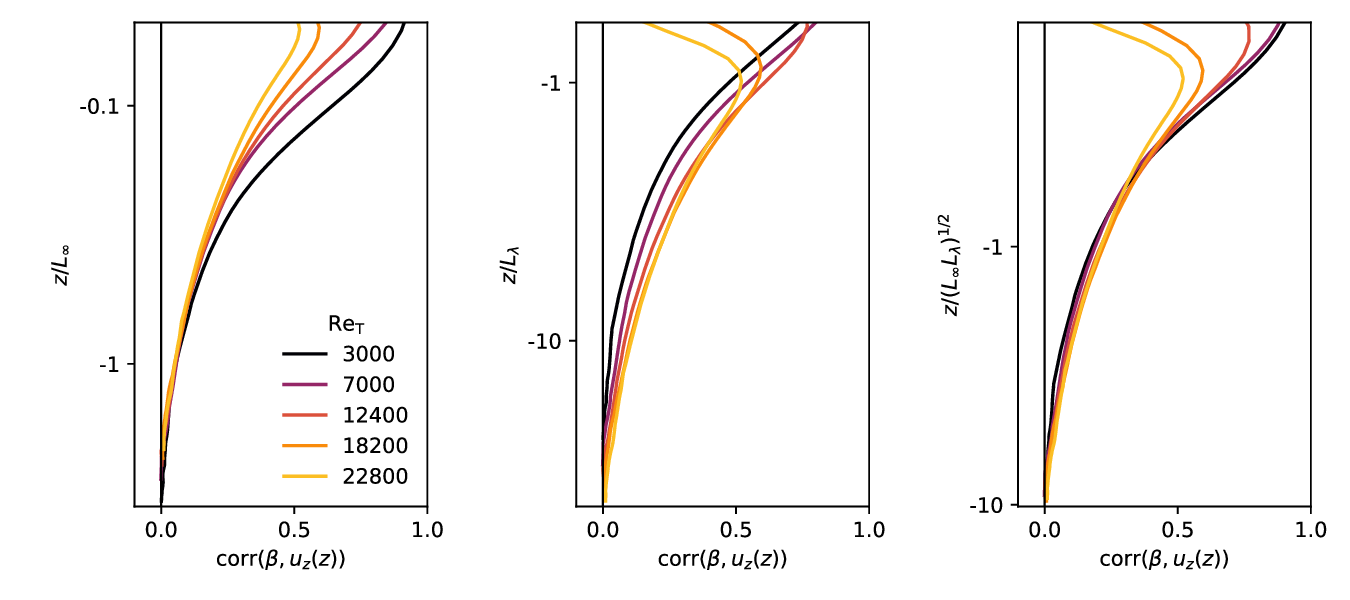}\put(6.5,42){(a)}\put(39,42){(b)}\put(72,42){(c)}\end{overpic}}
  \caption{Correlations between the surface divergence and sub-surface velocity at a given depth, with the depth normalized by (a) the far-field integral scale, (b) the far-field Taylor scale, and (c) a mixed length scale combining the two. Correlations less than 1 at the surface result from a combination of limited resolution near the surface, experimental error, and finite Froude and Weber number effects.}
\label{fig:beta_uz_corrs_vsz}
\end{figure}

Because the turbulent scales change throughout the source layer, there is no immediately apparent metric to characterize the size of upwelling and downwelling structures. As they involve vertical velocity fluctuations carrying fluid to or from the surface, however, the depth at which $C_{\beta, u_z}$ remains high embodies the reach of the surface-bulk coupling. Figure \ref{fig:beta_uz_corrs_vsz} shows profiles of $C_{\beta, u_z}$ versus $z$ normalized by three different length scales: $L_\infty$, $L_\lambda$, and the mixed length scale $(L_\lambda L_\infty)^{1/2}$. The latter incorporates the correlation lengths of both $\partial u_z / \partial z$ and $u_z$, yielding the best collapse of the data in the source layer (below the near-surface layer affected by viscous effects and surface deformation). Therefore, we conclude that this mixed scale, which involves the characteristic scales of the surface and sub-surface motions, is a viable estimate of the vertical extent of \new{surface-attached} upwellings and downwellings over the wide range of considered $\ReT$. 

In previous numerical studies, the horizontal footprint of these structures appeared to be comparable to $L_\infty$ \citep{guo_interaction_2010,herlina_direct_2014}. One potential explanation for the discrepancy with our results is the \dann{disparate Reynolds numbers}: the simulations attained $\ReT$ one-to-two orders of magnitude smaller than in our experiments and thus yielded marginal \new{separation between the relevant scales}, as \new{the Taylor scale scales as $L_\infty/L_\lambda \propto \ReT^{1/2}$ and the mixed length scale scales as $L_\infty / (L_\lambda L_\infty)^{1/2} \propto \ReT^{1/4} $}.

\subsection{Contribution to the inter-scale energy transfer}

By virtue of their different magnitude and topology, upwellings and downwellings contribute differently to the transport of energy in space and across scales. This is explored by conditioning the statistics on the sign of $u_z$ rather than $\beta$, which allows us to compare the turbulence structure associated to upward and downward fluctuations throughout the source layer. We still refer to upwellings/downwellings, though we do not restrict the analysis to surface-attached structures.

\begin{figure}
  \centerline{\begin{overpic}[width=1\linewidth]{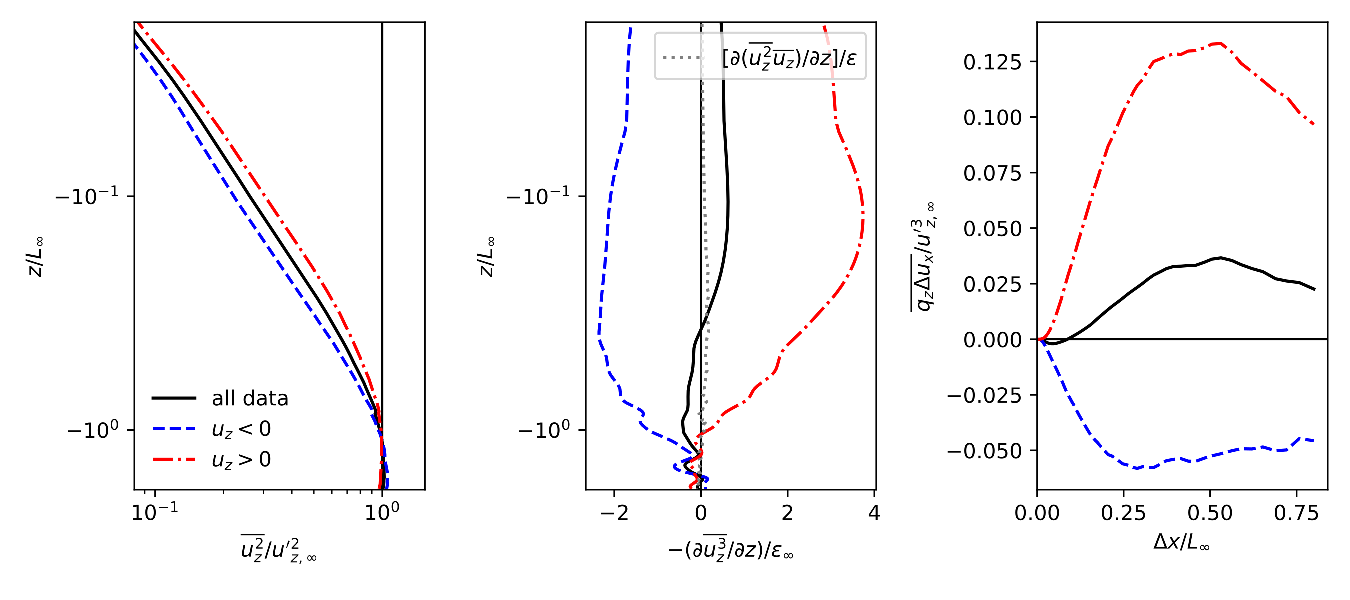}\put(6,42){(a)}\put(39,42){(b)}\put(72,42){(c)}\end{overpic}}
  \caption{The role of upwards and downwards motions in the budget of vertical TKE. (a) The vertical TKE profiles during each direction of motion. (b) The vertical \new{flux of vertical TKE} during each direction of motion. \new{The dashed gray line gives the flux due to the mean flow.} (c) The horizontal inter-scale transport of vertical TKE at a depth $z=-0.1 L_\infty$.}
\label{fig:conditional_transfer_z}
\end{figure}

Figure \ref{fig:conditional_transfer_z} (a) presents conditional profiles of the vertical component of TKE, indicating that upward motions carry stronger surface-normal fluctuations than downward ones: $\overline{u_z^2}^+ > \overline{u_z^2}^-$ (with superscripts indicating the sign of $u_z$). This is consistent with simulations by \citet{guo_interaction_2010}, who found the latter to have weaker surface-normal velocity than the former. The imbalance results from the spatial non-homogeneity in the source layer: downward motions carry fluid from the near-surface region where vertical TKE is lower, and vice versa for upward motions. This is reflected in the surface-normal transport of vertical TKE by the vertical fluctuations, $-\partial \overline{\dan{u_z^3}}/\partial z$ (figure \ref{fig:conditional_transfer_z} (b)). Its positive sign in the upper part of the source layer implies a net transport of turbulence towards the surface, as described in detailed by numerical simulations \citep{perot_shear-free_1995,walker_shear-free_1996,calmet_statistical_2003}. The net vertical transport results from opposite contributions (from upwellings and downwellings) of comparable magnitude, with upward motions prevailing especially at depths $O(0.1 L_\infty)$. This net transport has been shown to feed the net inter-component transport from vertical to horizontal energy \citep{walker_shear-free_1996}. By comparison, the net flux of vertical TKE by the small mean flow, $- \partial (\overline{u_z^2} \overline{u_z}) / \partial z$, is negligible. 

The differing behavior of downwellings and upwellings is connected to the decreased correlation between $q_z$ and $\Delta u_x$ near the surface, which was shown in section \ref{sec:inter-scale_energy_transfer} to determine the reduced cascade of vertical TKE. Specifically, as illustrated in figure \ref{fig:beta_uz_spatialcorr_bywelling}, downwellings and upwellings produce horizontal compression ($\overline{ \Delta u_x}^- < 0$) and stretching ($\overline{ \Delta u_x}^+ >0$) along the surface, respectively. The resulting inter-scale transfers of vertical TKE are displayed in figure \ref{fig:conditional_transfer_z} (c) for the representative depth $z/L_\infty=-0.1$: downwellings compress energy to smaller scales, while upwellings extend energy to larger horizontal scales. Due to the energetic imbalance shown in figure \ref{fig:conditional_transfer_z} (a), energetic extensions during upwellings are more effective than compressions during downwellings, ultimately resulting in the inverse cascade of vertical TKE.

\begin{figure}
  \centerline{\begin{overpic}[width=1\linewidth,grid=False]{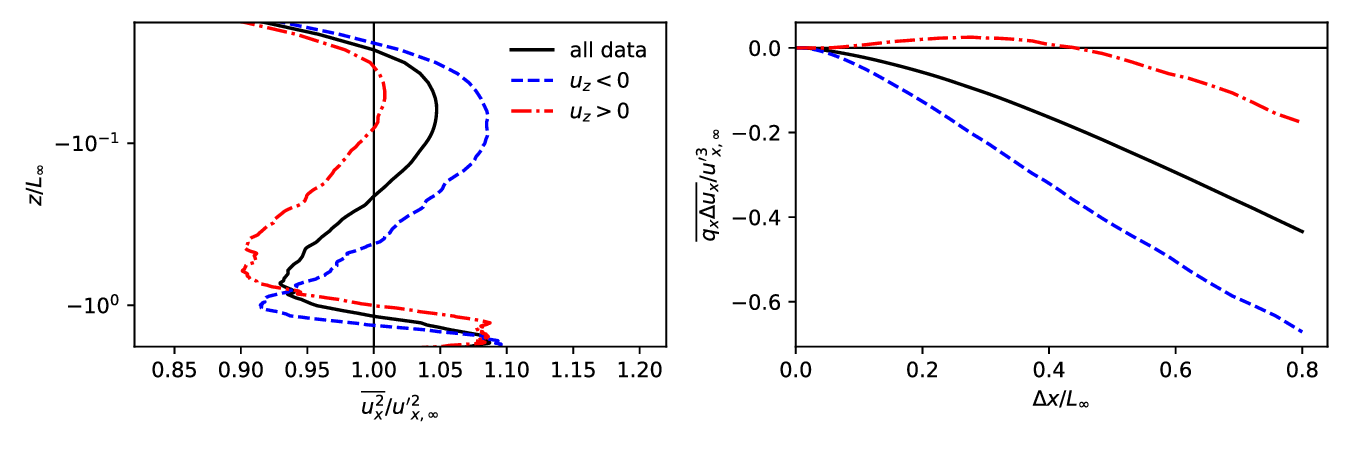}\put(6,32){(a)}\put(56,32){(b)}\end{overpic}}
  \caption{Characteristics of horizontal TKE during periods of downwards (blue) and upwards (red) motion. (a) The horizontal TKE during both types of motion. (b) The inter-scale horizontal transfer of horizontal TKE at $z=-0.1 L_\infty$, evidencing an inverse cascade during upwards motions. }
\label{fig:conditional_transfer_x}
\end{figure}

While upward motions contain a larger amount of vertical TKE compared to downward ones, the opposite is true for horizontal TKE: $\overline{u_x^2}^+ < \overline{u_x^2}^-$, as shown in figure \ref{fig:conditional_transfer_x} (a). Indeed, downward motions near the surface carry fluid from layers rich in horizontal energy, especially at the large scales, as described in section \ref{sec:horizontal_flucs}. Moreover, in keeping with the flow topology displayed in figure \ref{fig:beta_uz_spatialcorr_bywelling}, the horizontal TKE tends to be transferred to larger and smaller scales during upwellings and downwellings, respectively (figure \ref{fig:conditional_transfer_x} (b)). While shedding light on the role each type of motion plays in transferring horizontal TKE between scales, the present analysis does not fully explain the reduced correlation between $q_x$ and $\Delta u_x$ shown in figure \ref{fig:contributions_to_interscale_hinderance} (c), motivating future work.

\section{Conclusions}
\label{sec:conclusions}

We have investigated the influence a free surface exerts on the turbulence underneath, using a large zero-mean-flow water tank in which homogeneous turbulence of controllable intensity is forced. Several specific features of the present setup distinguish it from installations used in past studies, making it especially suitable for studying the problem. The turbulence in the bulk is homogeneous over a region much larger than the integral scale and has negligible mean velocity gradients; therefore, the finite size of the tank does not significantly influence the dynamics. The range of explored Reynolds numbers, up to $\ReT = 22800$ and $\mathrm{Re}_\lambda = 590$, allows for the development of an inertial range, with substantial separation between the integral and dissipative lengths of the system, $L_\infty / l_\mathrm{K} = O(10^3)$ and $L_\infty / \delta_\mathrm{v} = O(10^2)$. This is essential for establishing the power-law scalings predicted by \citet{kolmogorov_local_1941} and \citet{hunt_free-stream_1978}. Moreover, the forcing is applied homogeneously in depth up to less than one integral scale from the surface. This limits the spatial decay of TKE while maintaining weak surface deformation, with wave amplitudes of the order of the viscous layer thickness. This has enabled us to address several open questions, reaching the following conclusions.

In the source layer, both magnitude and length scale associated to the vertical TKE decrease approaching the surface, in line with the RDT predictions by \citet{hunt_free-stream_1978}. For most observables, the quantitative agreement with their theory systematically improves increasing $\ReT$. This is consistent with the analysis of \citet{magnaudet_high-reynolds-number_2003} who showed how nonlinear effects from the large-scale distortion by the surface (neglected in \new{the original analysis}) vanish in the high-$\ReT$ limit. The blockage effect is clearly demonstrated by the energy distribution across spatial scales: the surface limits the vertical fluctuations of eddies larger than the depth at which they are located. The increase of horizontal TKE predicted by RDT is visible only at the higher turbulence intensity, $\ReT > \sim 10000$, whereas for weaker forcing the effect is mild and thus obscured by spatial inhomogeneities. The level of forcing at which the horizontal TKE enhancement emerges is expected to depend on the specific system. Overall, our results indicate that differences in $\ReT$ and forcing schemes were the likely cause of discrepancy between previous studies.

The growth of horizontal energy in the source layer is concentrated at the large scales, specifically those for which the vertical energy is suppressed. This results in a strong enlargement of the integral scales of horizontal fluctuations, opposite to the RDT prediction. Such an accumulation of energy at the large scales is interpreted as the consequence of a hindered TKE cascade. The latter is demonstrated in the framework of the generalized Kármán–Howarth equation, specifically focusing on the inter-scale energy transfer across horizontal scales. The proximity to the surface inhibits the forward cascade of horizontal TKE, and even causes an inverse cascade of vertical TKE. This behaviour is rooted in a loss of correlation between energetic motions and compressive states of the flow. Such correlation is a hallmark of three-dimensional homogeneous turbulence, associated to the prevalence of vortex stretching and strain self-amplification and classically signalled by the negative skewness of the longitudinal velocity gradients \citep{davidson_turbulence_2004,carbone_is_2020,johnson_role_2021}. Near the surface, the extension/compression of velocity differences is radically altered by the upwelling and downwelling structures populating the near-surface region.

To analyse the effect of upwellings and downwellings on TKE transport, we have conditioned our data on the sign of the surface divergence and sub-surface velocity. Leveraging the scale separation achieved in our setup, we find that the vertical extent of up- and downwellings lies between the integral and the Taylor micro-scale, being $O((L_\lambda L_\infty)^{1/2}) \sim \ReT^{-1/4} L_\infty$. While a firm theoretical underpinning for such scaling is not available, a mixed length is consistent with the involvement of both energetic eddies (carrying fluid up the source layer) and velocity gradients (related to the surface divergence). Statistically, we find \new{upwellings to be more energetic}, determining the net flux of vertical TKE towards the surface. Downwellings, on the other hand, carry stronger horizontal TKE. These imbalances are connected to the opposite contribution of both types of motions to the inter-scale flux of energy: upwellings carry fluid parcels towards the surface and stretch them horizontally along it, while downwellings compress and carry them towards the bulk. Therefore, it is during downwellings that surface-attached vortices can stretch \citep{shen_surface_1999}, which is crucial for transferring horizontal energy to smaller scales \citep{davidson_turbulence_2004,johnson_role_2021}. 

The nature of the energy cascade in the vicinity of and along the free surface have been much debated, with several studies presenting evidence of a quasi-2D turbulent dynamics \citep{pan_numerical_1995,perot_shear-free_1995,sarpkaya_vorticity_1996,lovecchio_upscale_2015}, and others emphasizing the fundamentally 3D character of the flow \citep{walker_shear-free_1996,shen_surface_1999,guo_interaction_2010}. The present investigation represents a step to reconcile those views, as it highlights how upwellings and downwellings are not only chiefly responsible for the spatial transfer of energy, but also for the inter-scale flux at the surface. Energetic imbalances between upwellings and downwellings impact the amount of energy each type of motion extends or compresses to different scales. In the aggregate, \new{we find that} the near-surface structures modify the turbulence in such a way that the correlation between compressive and energetic structures is reduced, hindering the down-scale cascade of TKE. 

The present configuration in which turbulence is forced throughout the fluid volume is of high practical relevance; e.g., for shallow rivers and oceanic fronts, in which near-surface processes generate and sustain energy fluctuations \citep{nezu_turbulence_1993,rowinski_turbulence_2015,dasaro_enhanced_2011,taylor_submesoscale_2023}. Other common systems, however, involve turbulence generated at depth, diffusing towards the surface before feeling its influence. The effect of the \dann{distance between the \new{turbulence generation} region and the surface} has not been systematically assessed, and research is warranted on this point to identify mechanisms with a maximum degree of generality. 

Other notable aspects that are outside the scope of the present work deserve attention. In particular, the essentially non-homogeneous and anisotropic character of near-surface turbulence implies that 3D measurements are required to close the inter-scale energy budget. This is highly challenging as the Kolmogorov and integral scales need to be simultaneously resolved; it can be achieved, however, with advanced imaging approaches \citep{knutsen_inter-scale_2020}. Moreover, surface contamination may play a key role in the coupling of the sub-surface velocity to the surface divergence: Marangoni stresses induced by surfactant concentration gradients alter the structure of the divergence field \citep{mckenna_role_2004,shen_effect_2004}. Dedicated experiments are required to reach a predictive understanding of such processes. \new{Further}, when the surface deformation becomes large, its dynamics are two-way coupled with the turbulence dynamics underneath \citep{brocchini_dynamics_2001,savelsberg_experiments_2009,smeltzer_experimental_2023}. Future measurements \new{involving more highly-deformed surfaces will} elucidate this interplay of surface energy, wave \new{kinetic} energy and turbulence energy.

\backsection[Supplementary data]{\label{SupMat}A video of the vorticity fields at each Reynolds number is available as supplementary material.}


\backsection[Funding]{Funding from the Swiss National Science Foundation (project \# 200021-207318) is gratefully acknowledged.}

\backsection[Declaration of interests]{The authors report no conflict of interest.}


\backsection[Author ORCIDs]{D. J. Ruth, https://orcid.org/0000-0002-3764-4227; F. Coletti, https://orcid.org/0000-0001-5344-2476}

\bibliographystyle{jfm}
\bibliography{zotero_turbfreesurfaces}

\end{document}